\documentclass[epj]{webofc}
\usepackage[utf8]{inputenc}
\usepackage[varg]{txfonts}   
\usepackage{booktabs}
\usepackage{xcolor}
\definecolor{darkred}{rgb}{0.4,0.0,0.0}
\definecolor{darkgreen}{rgb}{0.0,0.4,0.0}
\definecolor{darkblue}{rgb}{0.0,0.0,0.4}
\usepackage[bookmarks,linktocpage,colorlinks,
    linkcolor = darkred,
    urlcolor  = darkblue,
    citecolor = darkgreen]{hyperref}
\usepackage{subfigure}
\usepackage{slashed}
\usepackage{caption}
\wocname{EPJ Web of Conferences}
\woctitle{Lattice2017}
\begin{document}
%
\selectlanguage{english}
\title{Nucleon axial coupling from Lattice QCD
}
\author{%
\firstname{Chia Cheng} \lastname{Chang}\inst{1}\thanks{Speaker, \email{chiachang@lbl.gov}}
\firstname{Amy} \lastname{Nicholson}\inst{2,1,3},
\firstname{Enrico}  \lastname{Rinaldi}\inst{4,1},
\firstname{Evan}  \lastname{Berkowitz}\inst{6},
\firstname{Nicolas}  \lastname{Garron}\inst{7},
\firstname{David}  \lastname{Brantley}\inst{8,1},
\firstname{Henry}  \lastname{Monge-Camacho}\inst{8,1},
\firstname{Chris}  \lastname{Monahan}\inst{9},
\firstname{Chris}  \lastname{Bouchard}\inst{10,8},\\
\firstname{M.A.}  \lastname{Clark}\inst{11},
\firstname{B\'alint}  \lastname{Jo\'o}\inst{12},
\firstname{Thorsten}  \lastname{Kurth}\inst{1},
\firstname{Kostas}  \lastname{Orginos}\inst{8,12},\\
\firstname{Pavlos}  \lastname{Vranas}\inst{5,1}, \and 
\firstname{Andr\'e}  \lastname{Walker-Loud}\inst{1,5,8,12},
\fnsep
}
\institute{%
Lawrence Berkeley National Laboratory, Berkeley, CA
\and
University of California, Berkeley, CA
\and
University of North Carolina, Chapel Hill, NC
\and
RIKEN-BNL, Brookhaven National Laboratory, Upton, NY
\and
Lawrence Livermore National Laboratory, Livermore, CA
\and
Forschungszentrum J\"ulich, J\"ulich, Germany
\and
University of Liverpool, Liverpool, UK
\and
The College of William \& Mary, Williamsburg, VA
\and
Rutgers, Piscataway, NJ
\and
University of Glasgow, Glasgow, UK
\and
NVIDIA Corporation, Santa Clara, CA
\and
Thomas Jefferson National Accelerator Facility, Newport News, VA
}
\abstract{%
We present state-of-the-art results from a lattice QCD calculation of the nucleon axial coupling,
$g_A$, using M\"obius Domain-Wall fermions solved on the dynamical $N_f = 2 + 1 + 1$ HISQ
ensembles after they are smeared using the gradient-flow algorithm. Relevant three-point
correlation functions are calculated using a method inspired by the Feynman-Hellmann
theorem, and demonstrate significant improvement in signal for fixed stochastic samples.
The calculation is performed at five pion masses of $m_\pi\sim \{400, 350, 310, 220, 130\}$~MeV,
three lattice spacings of $a\sim\{0.15, 0.12, 0.09\}$~fm, and we do a dedicated volume study with
$m_\pi L\sim\{3.22, 4.29, 5.36\}$. Control over all relevant sources of systematic uncertainty are
demonstrated and quantified. We achieve a preliminary value of $g_A = 1.285(17)$, with a
relative uncertainty of 1.33\%.
}
\maketitle
\section{Introduction}\label{intro}
The nucleon axial coupling, $g_A$, parameterizes the interaction strength of the nucleon with the weak axial current in the Standard Model, and is one of the fundamental properties which governs nuclear physics. For example, the $\beta$-decay rate of a neutron to a proton is governed by the strength of $g_A$, and confonting the sub-percent determination of $g_A$ obtained from experiments~\cite{Olive:2016xmw} with precise determination of this coupling from Lattice QCD serves as a test for the weak structure of the Standard Model. Interpretation of potential signals from observation of neutrinoless double-beta decay in nuclear targets also depends on the strength of the axial coupling to the nucleon, and in general, multi-nucleon systems. Demonstration of control over all sources of systematic uncertainty related to single nucleon $g_A$ is needed before calculating precision multi-body corrections to $g_A$. Precision tests of neutrino properties also depend on $g_A$ as an input, and in general $F_A(Q^2)$, the nucleon axial form factor, which parameterizes the nucleon axial coupling as a function of momentum transfer. In particular, future experimental results from DUNE and T2HK will be very interesting in light of recent results from T2K which exclude no CP violation in the leptonic sector with 90\% confidence level~\cite{PhysRevLett.118.151801}, lending evidence to scenarios involving baryogenesis through leptogenesis. In the regime of quasi-elastic scattering, only coupling to a single nucleon is required, however determination of the shape of the form factor from lattice QCD at high precision is desired when interpreting next generation experimental results. Robust determination of the form factor at zero momentum transfer provides a cross check with experimental measurements.

The calculation of $g_A$ from lattice QCD has, however, proven to be notoriously challenging and results have historically been in tension with the value determined from experiment. Insight of previous works has let us identify two main challenges that needed to be overcome in order to obtain a precise and accurate value of $g_A$: 1) the nucleon suffers from an exponentially problematic signal-to-noise problem in comparison to pions, leading to large statistical uncertainties and poor estimation of systematics at moderate values of source-sink separation, and 2) the excited state nucleon structure is dense compared to pion quantities and objectively demonstrating control over excited state contributions is required to remove this leading systematic uncertainty.

In this work we present a calculation of the nucleon axial coupling, $g_A= 1.285(17)$, based on a new computational method inspired by the Feynman-Hellmann theorem which alleviates the above concerns. This result is commensurate with the value determined from experiment, $g_A^{\textrm{PDG}} = 1.2723(23)$, to a precision of $1.33\%$. We summarize our computational method in Sec.~\ref{sec2}, the lattice setup in Sec.~\ref{sec3}, the construction of correlation functions in Sec.~\ref{sec4.1}, the analysis of the correlation functions in Sec.~\ref{sec4}, our renormalization procedure in Sec.~\ref{sec5}, the physical point extrapolation in Sec.~\ref{sec6}, sources of systematic error in Sec.~\ref{sec7}, and conclude in Sec.~\ref{sec8}.

\section{A Feyman-Hellmann inspired method}
\label{sec2}
For previous calculations of $g_A$~\cite{Edwards:2005ym,Capitani:2012gj,Horsley:2013ayv,Bali:2014nma,Abdel-Rehim:2015owa,Bhattacharya:2016zcn}, and in general, for processes in which an external current mediates the interaction (this excludes, for example, neutral meson mixing~\cite{Bazavov:2016nty,Bazavov:2017weg}), the three-point correlation function is constructed using the \textit{fixed-sink} method. The fixed-point correlation function is constructed from a sequential propagator, in which a regular one-to-all propagator is inverted again at the sink, at fixed time location. As a result, multiple fixed-sink correlation functions need to be calculated in order to reveal the source-sink separation dependence of the three-point correlation function, adding linearly to the computational cost per sink location.

For this work we use a novel method which, using the same computational cost as a single fixed-sink time, gives access to the complete sink time dependence. Having the full time dependence of the three-point correlation function availed us to the use of both exponentially more precise data at small source-sink separation times, and the full functional dependence needed to control excited state effects, directly addressing the two main challenges encountered while calculating nucleon related quantities. Furthermore, while the fixed-sink method allows for reuse of computational resources in the calculation of different currents within the same state, our method instead will allow us to reuse the resources for the calculation of the same current within different states, such as $g_A$ for multi-nucleon states or the axial charge of hyperons.

\subsection{Feynman-Hellmann correlation function}
Here we summarize the Feynman-Hellmann theorem inspired method used to perform our calculation, which is discussed in detail in Ref.~\cite{Bouchard:2016heu}. The Feynman-Hellmann theorem relates matrix elements to variations in the spectrum~\cite{Guettinger1932, Pauli1933, Hellmann1933, PhysRev.56.340},
\begin{equation}
\frac{\partial E_n}{\partial \lambda} = \langle n | H_\lambda| n \rangle,
\end{equation}
where the Hamiltonian is given by $H=H_0 + \lambda H_\lambda$. We apply the Feynman-Hellmann theorem \textit{directly} on to the analogous lattice energy, the $\textit{effective mass}$,
\begin{equation}
m^{\textrm{eff}}(t) = \ln\left(\frac{C(t)}{C(t+1)}\right),
\end{equation}
yielding
\begin{equation}
\left.\frac{\partial m^{\textrm{eff}}_\lambda(t)}{\partial \lambda} \right|_{\lambda=0} = \left[ \frac{\partial_\lambda C_\lambda(t)}{C(t)} - \frac{\partial_\lambda C_\lambda(t+1)}{C(t+1)}\right]_{\lambda=0},
\label{eq:meff_derivative}
\end{equation}
where $C(t)$ is the two-point correlation function. In the literature, there exists similar methods which adopt the name of the \textit{Feynman-Hellmann method}~\cite{Chambers:2014qaa,Chambers:2015bka} or a \textit{background field method}~\cite{Savage:2016kon}. The Feynman-Hellmann method that we employ for this calculation differs from the cited methods in that similar methods numerically calculate the derivative of the two-point correlation function of Eq.~\ref{eq:meff_derivative}, while we algebraically resolve this derivative. Numerical evaluation of the two-point correlation function requires the introduction of a background field at different values of the coupling $\lambda$, and modifies the derivative in Eq.~\ref{eq:meff_derivative} as a finite difference at additional computational cost for each value of the coupling.

The algebraic derivative of the two-point correlation function may be derived from the QCD generating functional with an external source $\mathcal{J}(t) \equiv \int d^3x j(t,\vec{x})$ with coupling $\lambda$, and nucleon interpolating operators $\mathcal{N}$ such that
\begin{align}
\left.-\frac{\partial C_\lambda(t)}{\partial \lambda} \right|_{\lambda=0} = -C(t)\int dt^\prime \langle \Omega \left| \mathcal{J}(t^\prime)\right|\Omega\rangle + \int dt^\prime \langle \Omega |T \{\mathcal{N}(t)\mathcal{J}(t^\prime)\mathcal{N}^\dagger(0)\}|\Omega\rangle.
\label{eq:FH_correlator}
\end{align}
The first term contains the vacuum expectation value of the current and vanishes for the axial current. The second term is a time-ordered three-point function integrated over the current insertion time $t^\prime$ and yields four contributions: 1) current insertion between the source and sink which contains the signal of interest, 2) current insertion outside the source and sink, and 3) and 4) contact terms when the current is on the same time slice as the source or sink. Note that a direct application of the Feynman-Hellmann theorem on the effective mass yields contribution 1), which reproduces the \textit{summation method}~\cite{Capitani:2012gj} if the summation method is applied to the fixed-sink correlator for \textit{all} sink locations.

\subsection{Spectral decomposition}
The spectral decomposition of the derivative of the two-point correlator given in Eq.~(\ref{eq:FH_correlator}), including the undesired contributions is derived in detail in Sec.~II~A of Ref.~\cite{Bouchard:2016heu}. We reproduce here the relevant conclusions of that paper.

The expression for the spectral decomposition of the three-point correlation function generated from the Feynman-Hellmann method (Eq.~(\ref{eq:FH_correlator})), may be simplified to the following form,
\begin{align}
N(t)&=\sum_n \left[(t-1)z_n g_{nn} z_n^\dagger +d_n \right] e^{-E_nt}
	+\sum_{n \neq m} z_n g_{nm} z_m^\dagger \frac{e^{-E_nt}e^{\frac{\Delta_{nm}}{2}}-e^{-E_mt}e^{\frac{\Delta_{mn}}{2}}}{e^{\frac{\Delta_{mn}}{2}}-e^{\frac{\Delta_{nm}}{2}}}.
\label{eq:FH_spectral}
\end{align}
As a result of the sum over the current insertion, the $t$ in the above expression is the source-sink time separation. The first term in the square brackets that receives the $(t-1)$ temporal enhancement contains $g_{nm} = \langle n|\mathcal{J}|m \rangle/\sqrt{4E_nE_m}$, where $n=m=0$ is the bare nucleon axial coupling, and is the dominant signal in the entire correlator. The overlap factor has a factor of energy absorbed into the definition as well, $z_n=\langle\Omega|\mathcal{N}|n\rangle/\sqrt{2E_n}$. The rest of the sum over $n$ for the $(t-1)$ term, and the sum over $n\neq m$ contains the excited state contamination when the current insertion is between the source and sink. We see that the additional artifacts introduced by the Feynman-Hellmann method (when compared to the fixed-sink method), may be cleanly parameterized by one additional tower of parameters $d_n$. This is illuminated by introducing the substitutions performed to obtain Eq.~(\ref{eq:FH_spectral}),
\begin{align}
d_n\equiv & Z_n Z^\dagger_{J:n} + Z_{J:n}Z^\dagger_n + Z_n Z^\dagger_n \langle\Omega | \mathcal{J} | \Omega\rangle + \sum_j \frac{Z_n Z^\dagger_{nj} J_j^\dagger + J_jZ_{jn}Z^\dagger_n}{2E_j\left(e^{E_j}-1\right)}.
\label{eq:FH_artifact}
\end{align}
The first two terms of Eq.~(\ref{eq:FH_artifact}) are from contact terms, and give rise to an analogous overlap factor $Z_{J:n}$, where the interpolating operator, which still overlaps with the nucleon, consists of three quarks and a quark-antiquark pair. The last two contributions come from when the current is inserted after the nucleon annihilation operator, resulting in baryon-meson transition matrix elements, $Z_{jn}$, and overlap factors on mesonic states, $J_j$. In principle, these extra contributions are difficult to determine. However, the entirety of $d_n$ is time independent, and can be simply reparameterized. In addition, note that,
\begin{align}
N(1) = \sum_n d_n e^{-E_n}.
\end{align}
As a result, inspecting the correlator at $t=1$ yields an extremely good order of magnitude estimate for the values of $d_n$.

\section{Domain-wall fermions on gradient-flowed HISQ}
\label{sec3}

\begin{table}
	{\scriptsize
			\begin{tabular}{llrrcc|cccccccc}
				\hline\hline
				\multicolumn{6} {c|} {\bfseries HISQ gauge configuration parameters} & \multicolumn{8} {c} {\bfseries valence parameters}  \\
				abbr. & $N_{\textrm{cfg}}$ & volume & $\begin{matrix} \sim a \\ \textrm{[fm]} \end{matrix}$ & $\begin{matrix} \sim m_{\pi_5} \\ \textrm{[MeV]} \end{matrix}$ & $\sim m_{\pi_5} L$ & $N_{\textrm{src}}$ & $L_5/a$ & $aM_5$ & $b_5$ & $c_5$ & $am_l^{\textrm{val.}}$ & $\sigma_{\textrm{smr}}$ & $N_{\textrm{smr}}$ \\
				\hline
				a15m310  & 1960 & $16^3\times48$ & 0.15 &  310 & 3.8 & 24& 12 & 1.3 &   1.5 &   0.5 & 0.01580 & 4.2 &   60 \\
				a15m220  & 1000 & $24^3\times48$ & 0.15 & 220 & 4.0 & 12& 16 & 1.3 & 1.75 & 0.75 & 0.00712 & 4.5 &   60 \\
				a15m130 &  1000 & $32^3\times48$ & 0.15 & 130 & 3.2 &   5& 24 & 1.3 & 2.25 & 1.25 & 0.00216 & 4.5 &   60 \\
				\hline
				a12m400  & 1000 & $24^3\times64$ & 0.12 &  400 & 5.8 &   8&   8 & 1.2 & 1.25 & 0.25 & 0.02190 & 5.0 &   75 \\
				a12m350  & 1000 & $24^3\times64$ & 0.12 &  350 & 5.1 &   8&   8 & 1.2 & 1.25 & 0.25 & 0.01660 & 5.0 &   75 \\
				a12m310  & 1053 & $24^3\times64$ & 0.12 &  310 & 4.5 &   4&   8 & 1.2 & 1.25 & 0.25 & 0.01260 & 5.0 &   75 \\
				a12m220S& 1000 & $24^3\times64$ & 0.12 & 220 & 3.2 &   4& 12 & 1.2 &   1.5 &   0.5 & 0.00600 & 6.0 &   90 \\
				a12m220  & 1000 & $32^3\times64$ & 0.12 & 220 & 4.3 &   4& 12 & 1.2 &   1.5 &   0.5 & 0.00600 & 6.0 &   90 \\
				a12m220L& 1000 & $40^3\times64$ & 0.12 & 220 & 5.4 &   4& 12 & 1.2 &   1.5 &   0.5 & 0.00600 & 6.0 &   90 \\
				a12m130 &  1000 & $48^3\times64$ & 0.12 & 130 & 3.9 &   3& 20 & 1.2 &   2.0 &   1.0 & 0.00195 & 7.0 & 150\\
				\hline
				a09m310  &   784 & $32^3\times96$ & 0.09 & 310 & 4.5 &   8&   6 & 1.1 & 1.25 & 0.25 & 0.00951 & 7.5 & 167\\
				a09m220 &  1001 & $48^3\times96$ & 0.09 & 220 & 4.7 &   6&   8 & 1.1 & 1.25 & 0.25 & 0.00449 & 8.0 & 150\\
				\hline\hline
			\end{tabular}
		\caption{\label{tab:hisq} HISQ gauge configurations and valence sector parameters. The HISQ ensembles used in this work (e.g. a15m310 stands for the ensemble with $a\sim0.15$~fm and $m_\pi\sim310$~MeV) along with the number of configurations $N_{\textrm{cfg}}$, lattice volume, approximate lattice spacing $a$, approximate HISQ taste-5 pion mass, and approximate value of $m_{\pi,5} L$. The taste-5 pion is the one protected by $\gamma_5$-symmetry, and does not receive additive renormalizations. Values are obtained from Table I of Ref.~\cite{Bazavov:2012xda} with increased number of configurations.
		Mobius domain-wall propagators are generated at a number of sources per configuration $N_{\textrm{src}}$, with the fifth dimensional extent $L_5/a$, such that $m_{\textrm{res}}$ is minimized at $aM_5$, with the Mobius kernel defined by $b_5$ and  $c_5$, and valence light-quark masses $am_l^{\textrm{val.}}$. We also list the width $\sigma_{\textrm{smr}}$ and iteration count $N_{\textrm{smr}}$ of the \texttt{SHELL\_SOURCE} and the \texttt{GAUGE\_INV\_GAUSSIAN} smearing algorithm in \texttt{Chroma}.}
	}
\end{table}

The calculation is performed on the MILC collaboration's 2+1+1 flavor Highly-Improved Staggered Quark (HISQ) ensemble~\cite{Bazavov:2010ru,Bazavov:2012xda}. The MILC ensembles are the only set of publicly available gauge configurations that allow for control over the continuum limit, infinite volume, and physical pion mass extrapolation. The MILC configurations used in this project span three lattice spacings, $a\sim\{0.15,0.12,0.09\}$~fm allowing for control over $a^2$ discretization corrections. Three pion masses are generated by the MILC collaboration, $m_\pi\sim \{130,220,310\}$~MeV, while additional heavier pion masses of $m_\pi\sim \{350,400\}$~MeV are generated by the CalLat collaboration to control interpolation to the physical pion mass. A volume study is performed at $a\sim0.12$~fm, $m_\pi\sim220$~MeV, where all input parameters are held fixed at three spatial volumes $m_\pi L\sim \{3.2,4.3,5.4\}$. Formally the HISQ action has leading discretization errors starting at $\mathrm{O}(\alpha_S a^2, a^4)$, however improved link-smearing greatly suppresses taste-changing interactions, leading to numerically smaller coefficients. The gluons are described by the tadpole-improved~\cite{PhysRevD.48.2250}, one-loop Symanzik gauge action~\cite{Alford199587} with leading discretization errors starting at $\mathrm{O}(\alpha_S^2 a^2, a^4)$.

The valence quarks are calculated using the M\"obius domain-wall action~\cite{Brower:2004xi,Brower:2005qw,Brower:2012vk}. The mixed action set up is chosen because of the good chiral properties of the domain-wall action, which significantly suppresses chiral symmetry breaking and discretization effects from using a staggered sea. The M\"obius kernel allows for moderate values of the fifth dimensional extent $L_5$ to be chosen while still keeping the residual chiral symmetry breaking to less than 10\% of the input light quark mass. Due to the very small residual chiral symmetry breaking, numerically, the domain-wall action has discretization errors starting at $\mathrm{O}(a^2,\alpha_S a^2)$.  The valence pseudoscalar masses are tuned to within 2\% of the HISQ masses, yielding a unitary theory in the continuum limit.

\begin{figure}
	\includegraphics[width=0.5\columnwidth]{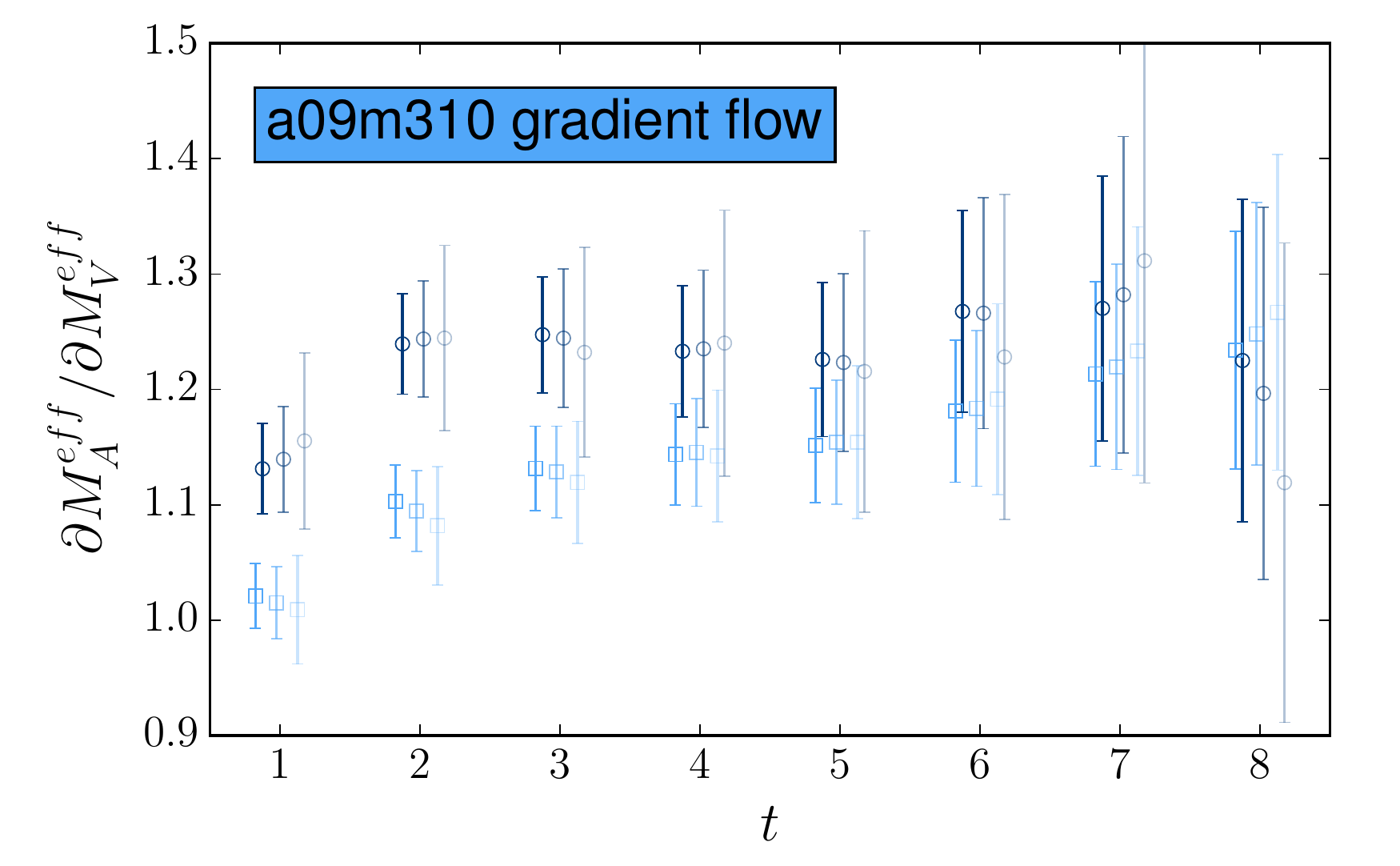}
	\includegraphics[width=0.5\columnwidth]{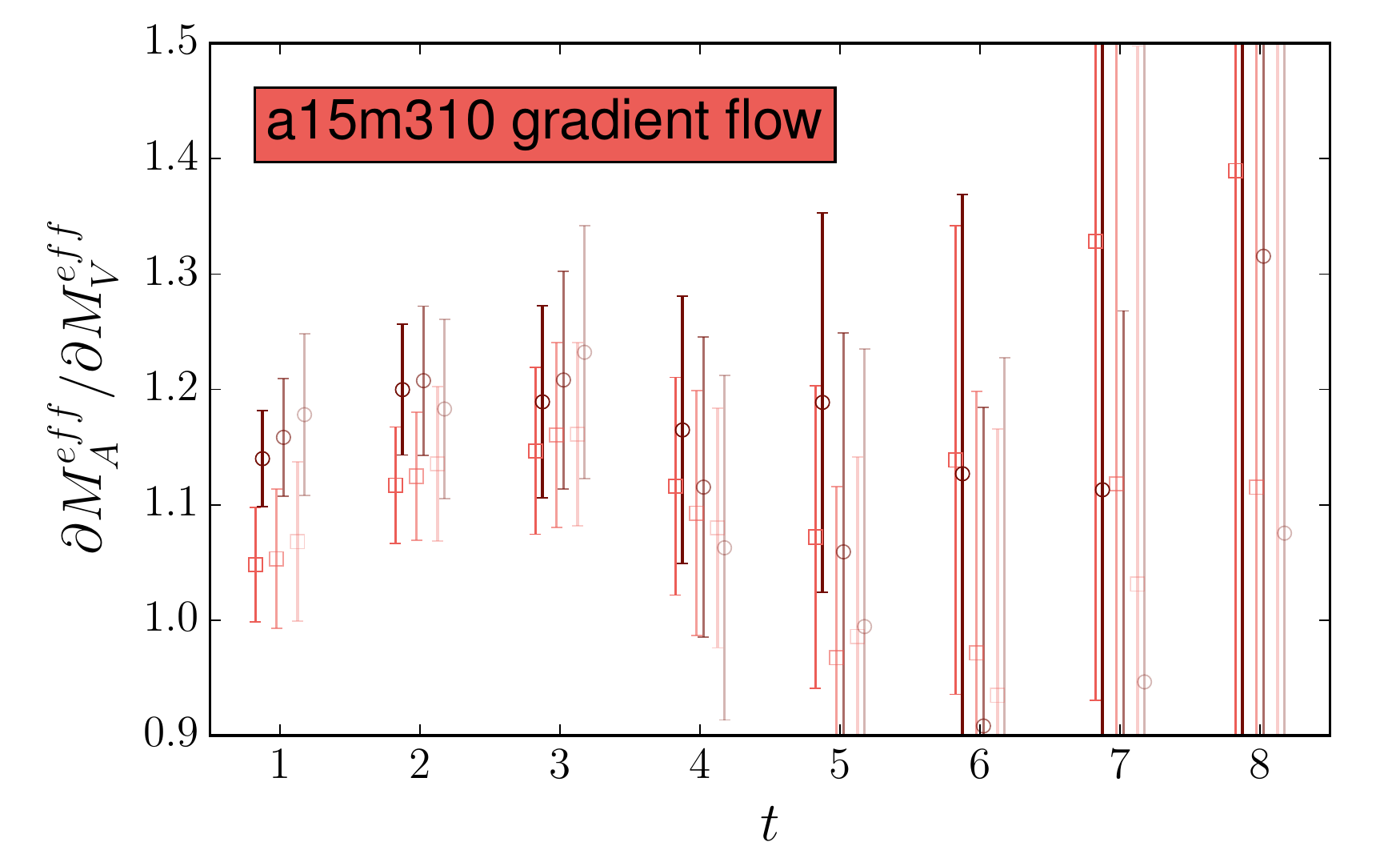}
	\caption{\label{fig:gAflow} The ratio of $g_A/g_V$ as a function of source-sink separation time $t$ with smeared (circle) and point (square) sink operators. At each time slice, from left to right, the ratio is evaluated at $t_{gf}=\{0.2,0.6,1.0\}$. Flow time dependence is plotted for the (Left) a09m310 and (Right) a15m310 ensembles.}
\end{figure}

We add an additional layer of gauge field smearing with the 4-dimensional Wilson flow procedure with a dimensionless flow time of $t_{gf}=1.0$~\cite{Luscher:2010iy,Lohmayer:2011si}. The new scale introduced by the gradient flow is approximately $l_{gf}\sim\sqrt{8t_{gf}}a$. We observe that the gradient flowed ensembles are more continuum-like, and are stochastically less noisy, because the new scale decouples UV contamination from the correlator. In addition, we have evidence supporting the fact that our action is not over smeared. For the axial coupling, we observe that the ratio of $g_A/g_V$ is flow time independent, as shown in Fig.~\ref{fig:gAflow}, demonstrating that the continuum extrapolation is flow time independent. Additionally, we performed a complete flow time study on $F_K/F_\pi$  calculated at $t_{gf}=\{0.2,0.4,0,6,0.8,1.0\}$. The ratio of pseudoscalar decay constants after chiral-continuum extrapolation is demonstrably flow time independent, and consistent with the FLAG average~\cite{Aoki:2016frl}.

Table~\ref{tab:hisq} lists parameters pertinent to the gauge configurations, and valence sector. A complete discussion of flow time dependence, and an in-depth review for this mixed action set up may be found at Ref.~\cite{Berkowitz:2017opd}.

\section{Lattice correlation functions}
\label{sec4.1}
Standard nucleon interpolating operators are used to when constructing the correlation functions~\cite{Basak:2005aq,Basak:2005ir},
\begin{align}
\bar{\mathcal{N}}_{\gamma^\prime}(x)=&\epsilon_{i^\prime,j^\prime,k^\prime}\left(\bar{u}^{i^\prime}_{\alpha^\prime}(x)\Gamma^{\textrm{src}}_{\alpha^\prime \beta^\prime}\bar{d}^{j^\prime}_{\beta^\prime}(x)\right)P^{\textrm{src}}_{\gamma^\prime \rho^\prime}\bar{u}^{k^\prime}_{\rho^\prime}(x),\nonumber\\
\mathcal{N}_{\gamma}(x)=&-\epsilon_{ijk}\left(u^i_\alpha(y)\Gamma^{\textrm{snk}}_{\alpha\beta}d^j_\beta(y)\right)P^{\textrm{snk}}_{\gamma\rho}u^k_\rho(y),\end{align}
where $u(x)$ and $d(x)$ are up- and down-quark field operators at $x$, while the $\Gamma$ and $P$ are spin projection operators. We double statistics by constructing the spin averaged correlation functions. The local axial and vector operator is used to construct the derivative correlator in Eq.~(\ref{eq:FH_correlator}), from which we obtain the bare values of $g_A$ and $g_V$,
\begin{align}
\mathcal{J}_{A_3}(x) = & \bar{q}_{\alpha}^i(x)\gamma_{3\alpha\beta}\gamma_{5\beta\gamma}\tau_3^{ij} q_{\gamma}^j(x),\label{eq:current_A}\\
\mathcal{J}_{V_4}(x) = & \bar{q}_{\alpha}^i(x)\gamma_{4\alpha\beta} \tau_3^{ij} q^j_\beta(x).
\label{eq:current_V}
\end{align}
The Pauli $\tau_3$ matrix acts on isospin space, and projects the isovector combination, yielding $g_A$ relevant to nuclear beta-decay. A detailed discussion on the construction of the lattice correlation function, and their associated Wick contractions are presented in Sec.~II~B of Ref.~\cite{Bouchard:2016heu}.

Multiple sources per configuration are created in order to increase our statistical sampling, as reported in the first column under \textit{valence parameters} of Table~\ref{tab:hisq}. For each configuration, a random origin is chosen, $(x_0,y_0,z_0,t_0)$, along with its antipode in space, $(x_0,y_0,z_0,t_0)+L/2(1,1,1,0)$, on which evenly spaced time locations are chosen for the multiple sources. Once generated, the correlators are shifted to $t_0=0$ and averaged over different sources. We observe a reduction in statistical uncertainty inversely proportional to $\sqrt{N_\textrm{src}}$, indicating no correlation between different sources. We further double the statistics by combining the time-reversed negative-parity nucleon correlator with the forward-propagating positive-parity nucleon correlator.

Correlators are constructed using a \textit{gauge invariant Gaussian} smeared source~\cite{Edwards:2004sx} to increase the overlap with the ground state, and increase statistics by constructing and analyzing correlators with smeared and point sinks. The last two columns of Table~\ref{tab:hisq} under \textit{valence parameters} provide the smearing width and iteration count of the procedure.

\begin{figure}
	\includegraphics[width=0.5\columnwidth]{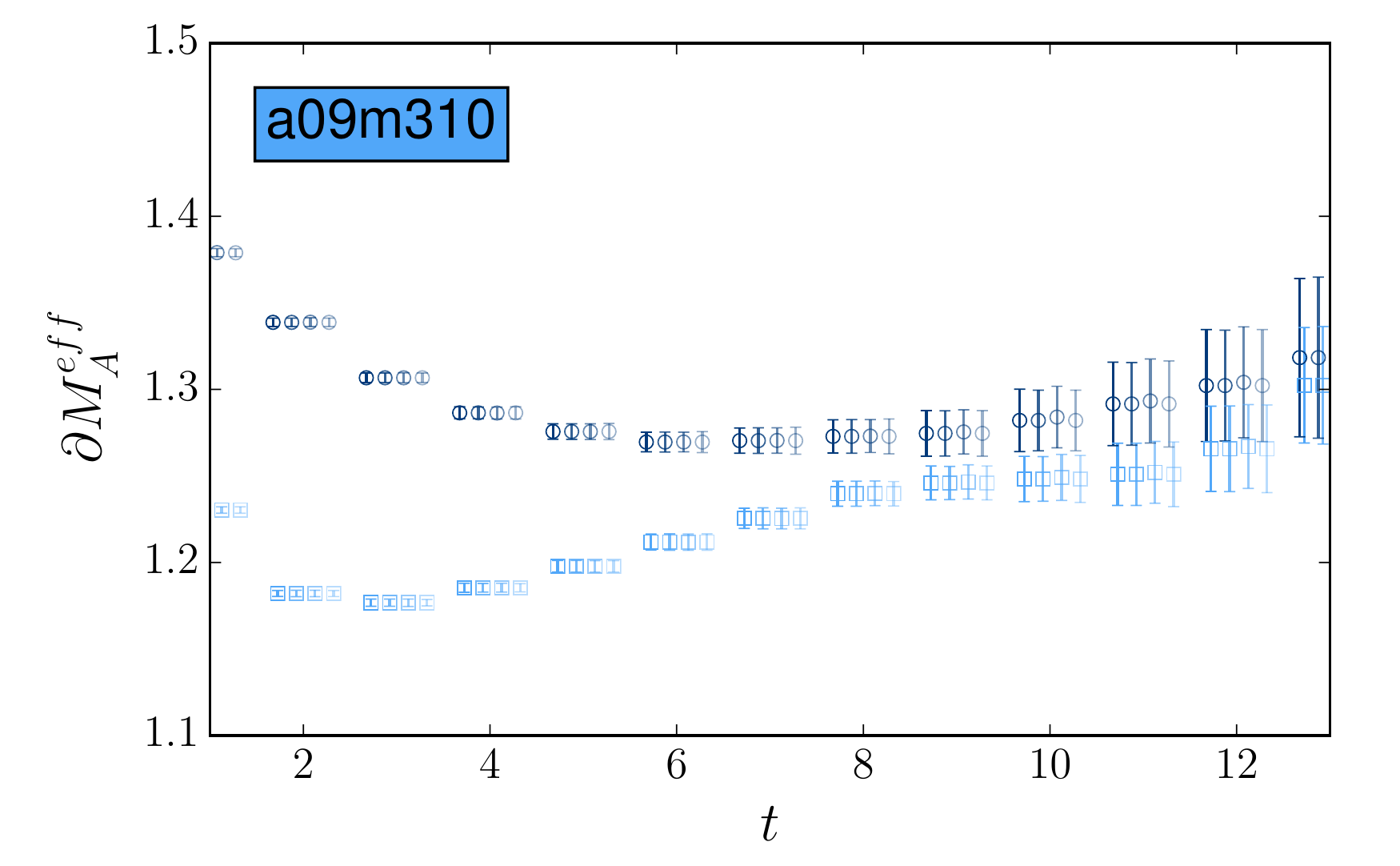}
	\includegraphics[width=0.5\columnwidth]{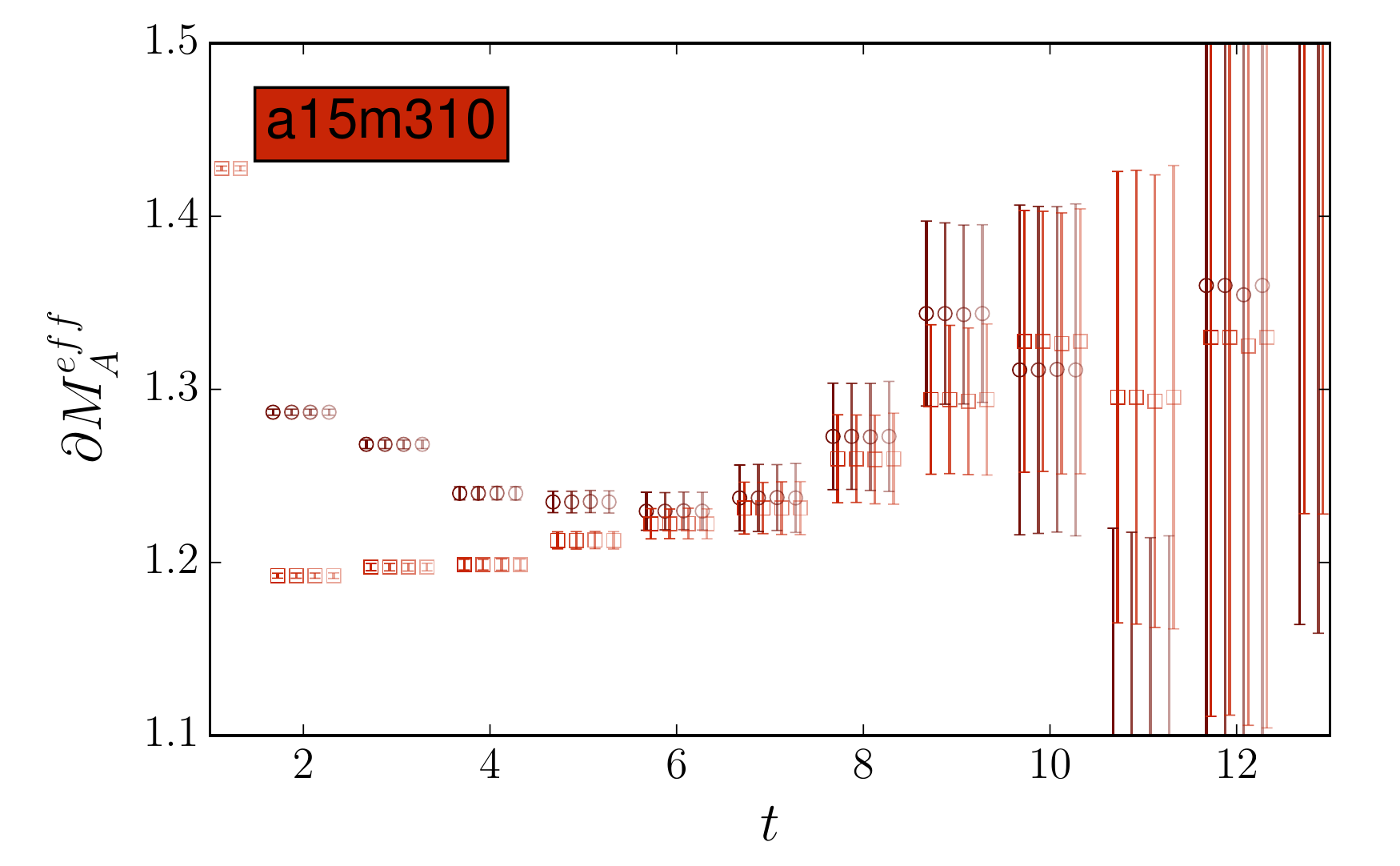}
	\caption{\label{fig:gAauto} Autocorrelation study for the $g_A$ correlation function. The derivative of the effective mass given by Eq.~(\ref{eq:meff_derivative}) is plotted as a function of source-sink separation time $t$ with smeared (circle) and point (square) sink operators. The standard error is plotted for each data point. In each time slice, the correlator is binned by every $\{1,2,3,4\}$ configurations. Bin size dependence is plotted for the (Left) a09m310 and (Right) a15m310 ensembles.}
\end{figure}

Finally, the dataset is free of autocorrelations as a function of Monte Carlo time as demonstrated by Fig.~\ref{fig:gAauto}. The error of the mean is stable as a function of bin size at all time slices for the a09m310 and a15m310 ensembles. The independence of bin size is representative of all the data. As a result, we do not bin any of our data.

\section{Correlation function analysis}
\label{sec4}
In the following section, we discuss our correlator fit procedure and provide evidence for full control over contamination of the ground state from contributions from excited states. A short discussion on the analysis of the pion mass and decay constants is also provided because a correlated analysis is performed in a later step when the chiral-continuum limit is taken. All correlator fits are performed using the Python library \texttt{lsqfit}~\cite{lsqfit}.

\subsection{Bare $g_A$ and $g_V$ fits}
The bare axial and vector couplings of each ensemble are extracted from a simultaneous and correlated fit to the two point correlation function and the effective mass derivatives given by Eq.~(\ref{eq:meff_derivative}) for the axial and vector current insertions, each with smeared- and point-sink operators. The six correlators are fit together because the overlap factors and the spectrum are shared. We first perform a two-state constrained fit in order to survey a wide range of source-sink separation times for all six correlators. The central values of the resulting posterior distributions are then used as initial guesses for the final unconstrained correlator fit. The initial step of constrained curve fitting does not affect the final result, but simply provides an efficient way to explore large parameter spaces. The initial guesses motivated by the mean of the posterior distributions also only decreases the number of steps required for convergence. The final unconstrained fit is performed using a two-state fit for all six correlators.

The preferred correlator fits are the ones which pass a set of stringent requirements, resulting in an objective selection process. We first assess the quality of all fits by considering only results with $p$-values greater than 0.05 in order to discriminate against fits of poor quality. The $p$-value in practice does an adequate job eliminating fits which are in great tension with data. This is usually seen when the fit is performed too aggressively at very small source-sink separations, where ``very small'' is of course, a relative statement. A good $p$-value, defined to be greater than 0.05,  however, is an inadequate condition for a good fit. To demonstrate control over excited state contamination, we vary the fit regions over different time separations, and demand that the preferred fit lie in the region of stability. The reasoning is that if the data is adequately described by our fit ansatz, which in the case of this calculation, we limit to the inclusion of only 2 states. This is because an unconstrained fit to more than two states leads to numerical instability. For the fit ansatz for $g_A$ and $g_V$, we directly implement the ratio given by Eq.~(\ref{eq:meff_derivative}) by constructing the difference in the ratio of Eq.~(\ref{eq:FH_spectral}) and the standard two-point spectral decomposition without performing any additional manipulation to the fit ansatz.

 \begin{figure}
	\includegraphics[width=0.5\columnwidth]{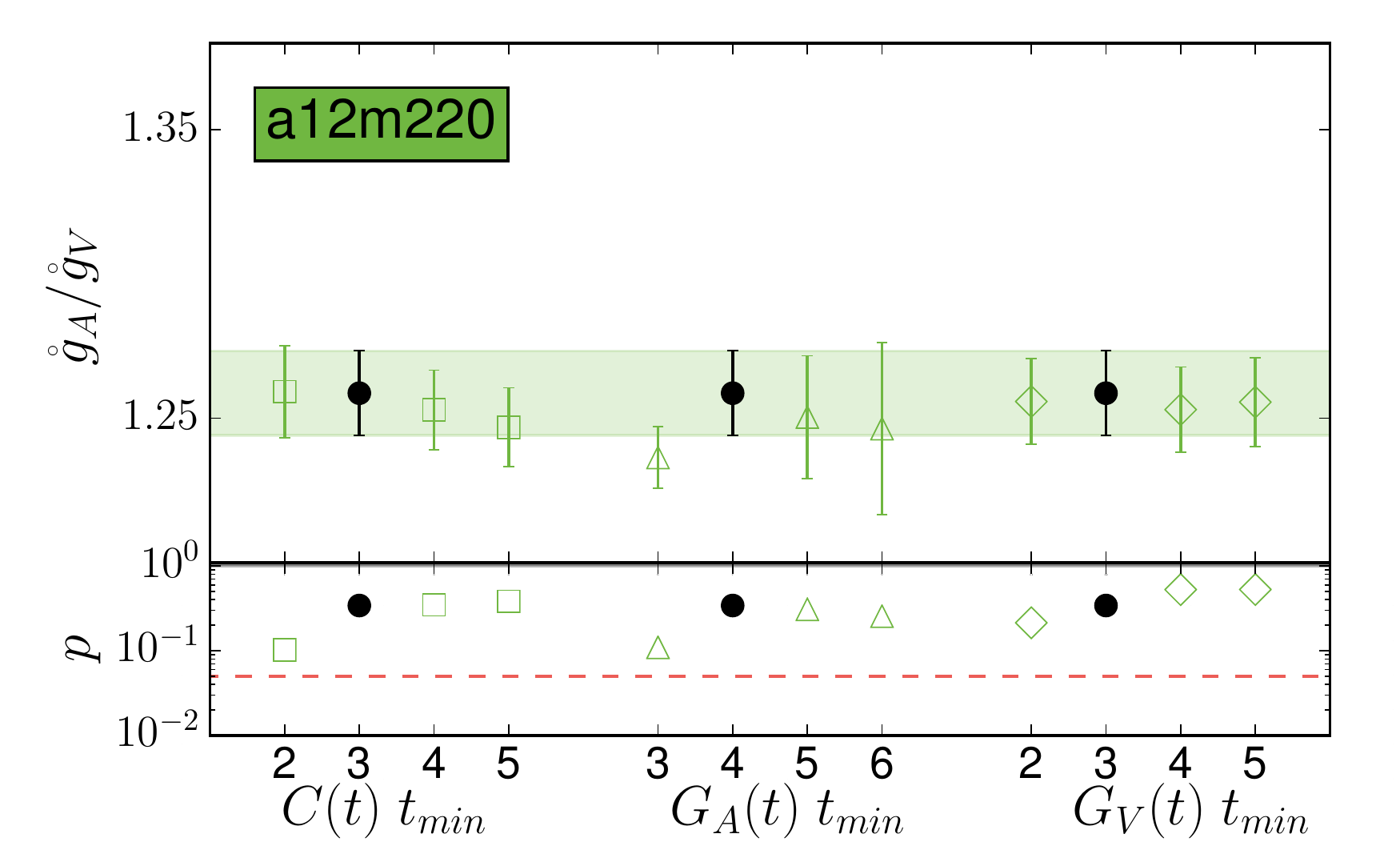}
	\includegraphics[width=0.5\columnwidth]{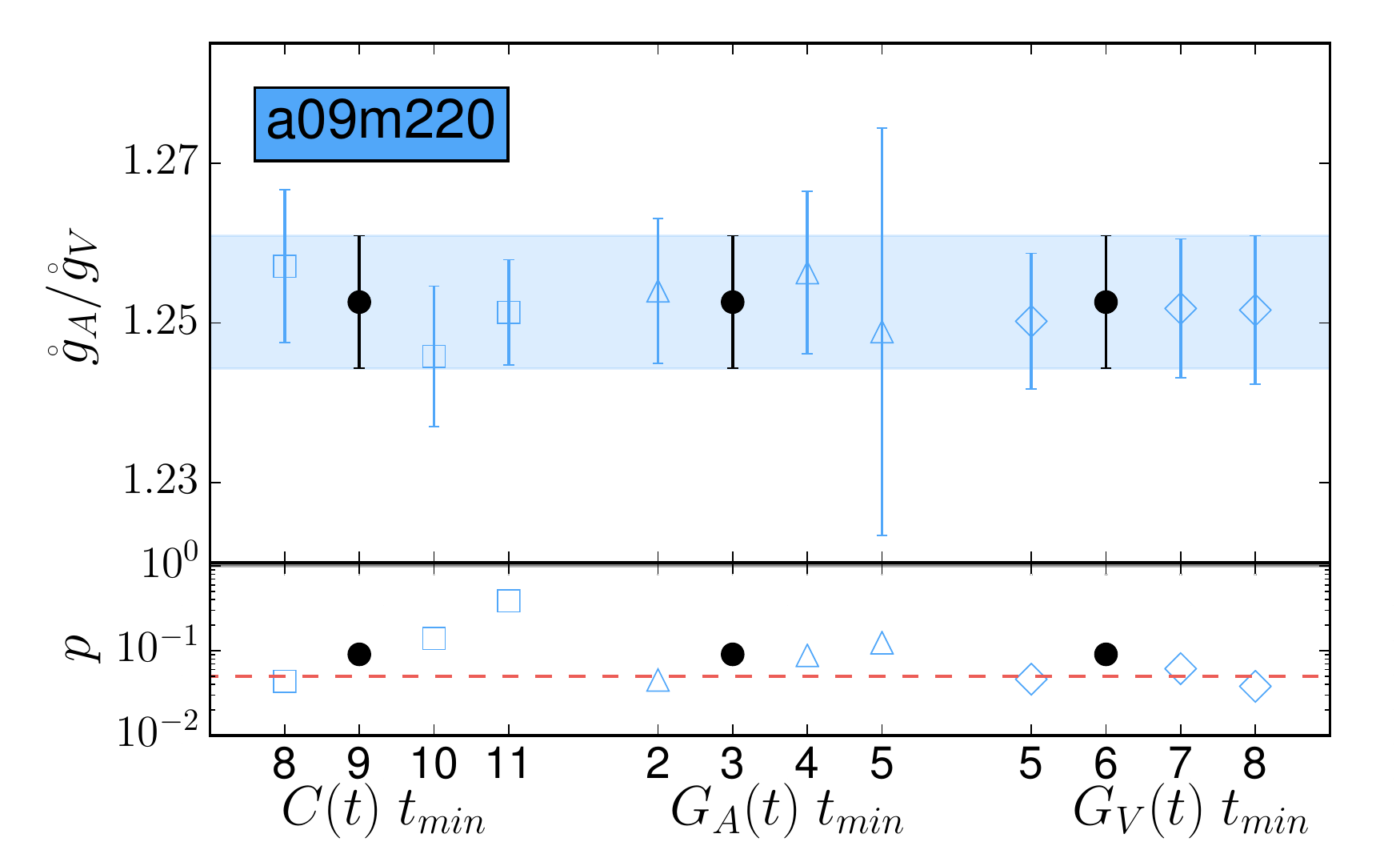}
	\caption{\label{fig:corrfit_example} Stability plot for the (Left) a12m220 and (Right) a09m220 ensembles. The uncertainty of the fit reflects the middle 68\% confidence interval. The ratio of the bare couplings $\mathring{g}_A/\mathring{g}_V$ under varying values of $t_{\textrm{min}}$ for the two point $C(t)$, $g_A$ correlator $G_A(t)$, and $g_V$ correlator $G_V(t)$. The preferred fit is highlighted in black. The horizontal band highlights the preferred fit and helps guide the eye. The corresponding $p$-value is plotted in the bottom panel, where the dashed red line is set at a $p$-value of 0.05, the minimum acceptable value. The values of $t_{\textrm{max}}$ is fixed to $\{15,10,14\}$ for $\{C(t), G_A(t), G_V(t)\}$.}
\end{figure}

In Fig.~\ref{fig:corrfit_example}, we show example fits for the a12m220 and a09m220 ensembles, which pass this test. The figure demonstrates that the preferred fit (in black) lies in the region of stability while the fit regions are varied over the two point $C(t)$, $g_A$ correlator $G_A(t)$, and $g_V$ correlator $G_V(t)$. The horizontal bands highlights the preferred fit, while the green (blue) points to the right lie consistently within the preferred fit. We also observe that by going to larger values of $G_A(t)~t_{\textrm{min}}$, the uncertainty on the ratio $\mathring{g}_A/\mathring{g}_V$ increases. This is because dropping more precise data leads to a more uncertain result. In general the uncertainty of the ratio is more sensitive to changes in the fit region to the $g_A$ correlator because the uncertainty on $\mathring{g}_V$ is approximately an order of magnitude smaller than $\mathring{g}_A$, while the two-point correlator only indirectly affects the ratio. However, dropping too much data from the two-point correlators (not shown) will eventually lead to catastrophic numerical instabilities because the overlap factors can only be disentangled from information provided by the two-point correlators. On the other hand, stability plots also reveal fits that are too aggressive, as shown by the fit at $G_A(t)~t_{\textrm{min}} = 3$ for the a12m220 ensemble. We observe that the more aggressive fit is more than one standard deviation lower than the preferred fit, lending evidence to residual excited state artifacts contaminating in the ground state parameter.  The different preferred fits also all pass the minimum $p$-value (indicated by the dashed red line), as demonstrated by the bottom portion of Fig.~\ref{fig:corrfit_example}.  Analysis of the a15m310 ensemble was performed in previous work~\cite{Bouchard:2016heu} under the Bayesian framework with up to 8 states and is consistent with the result presented here.

 \begin{figure}
	\includegraphics[width=0.5\columnwidth]{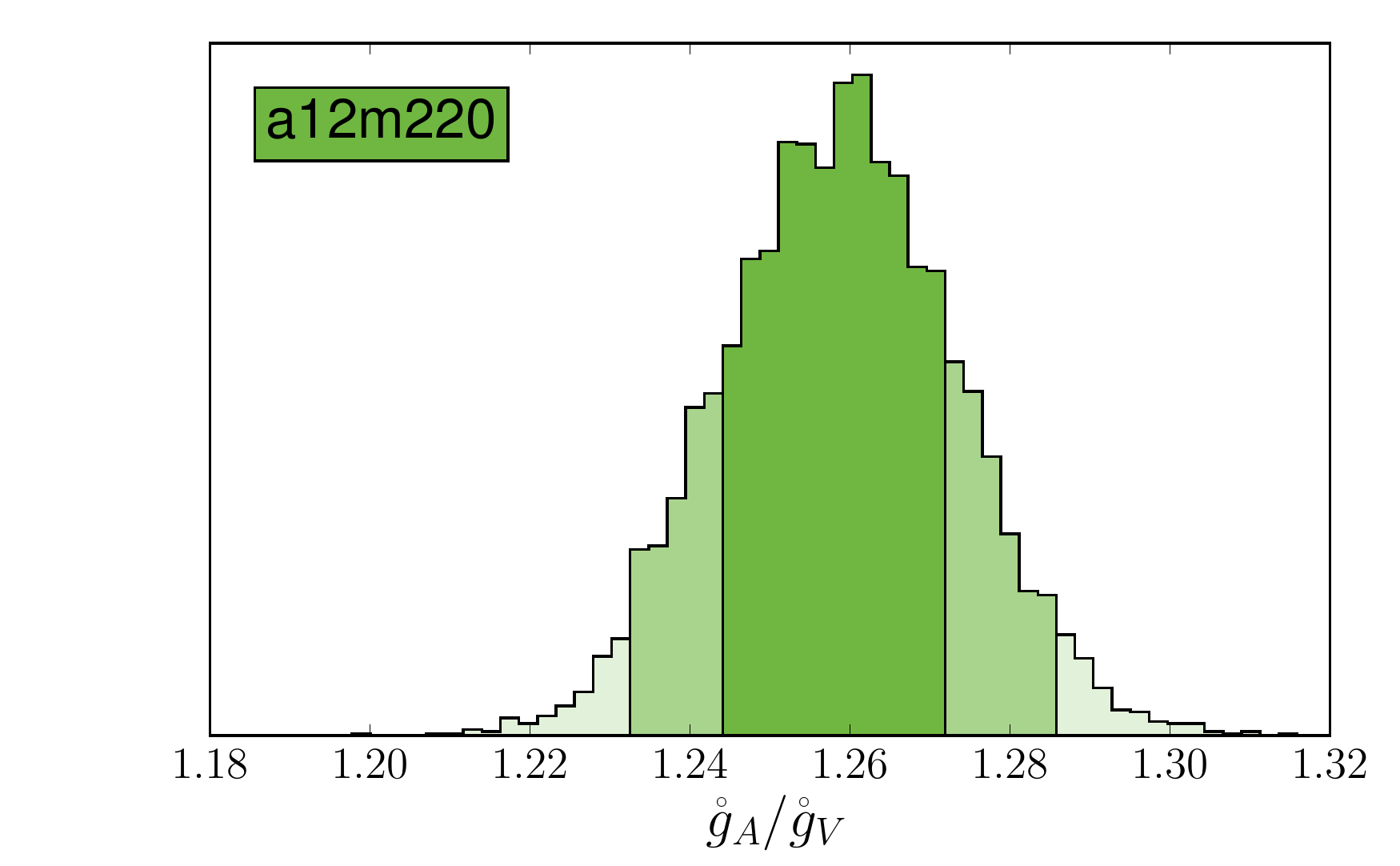}
	\includegraphics[width=0.5\columnwidth]{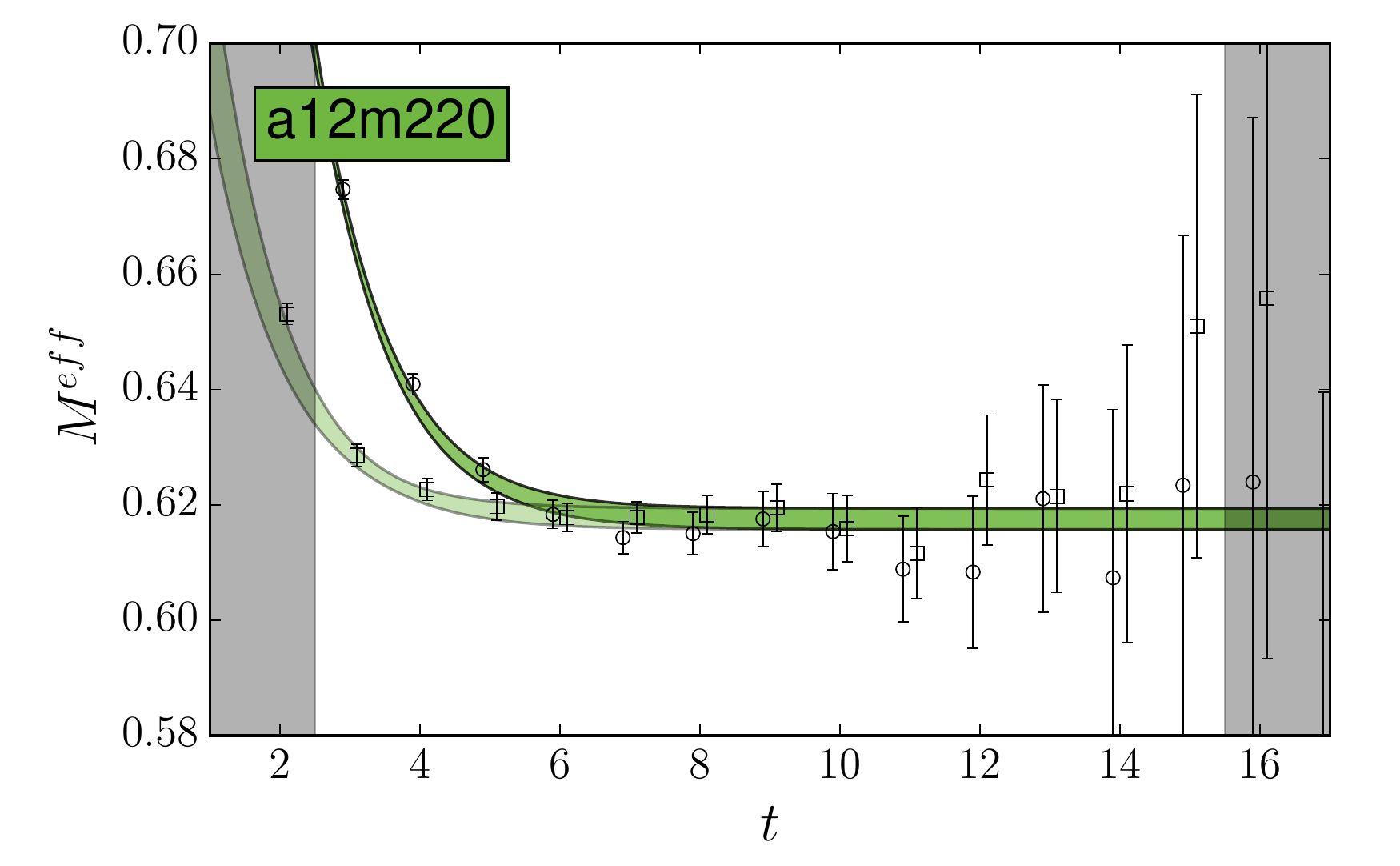}
	\includegraphics[width=0.5\columnwidth]{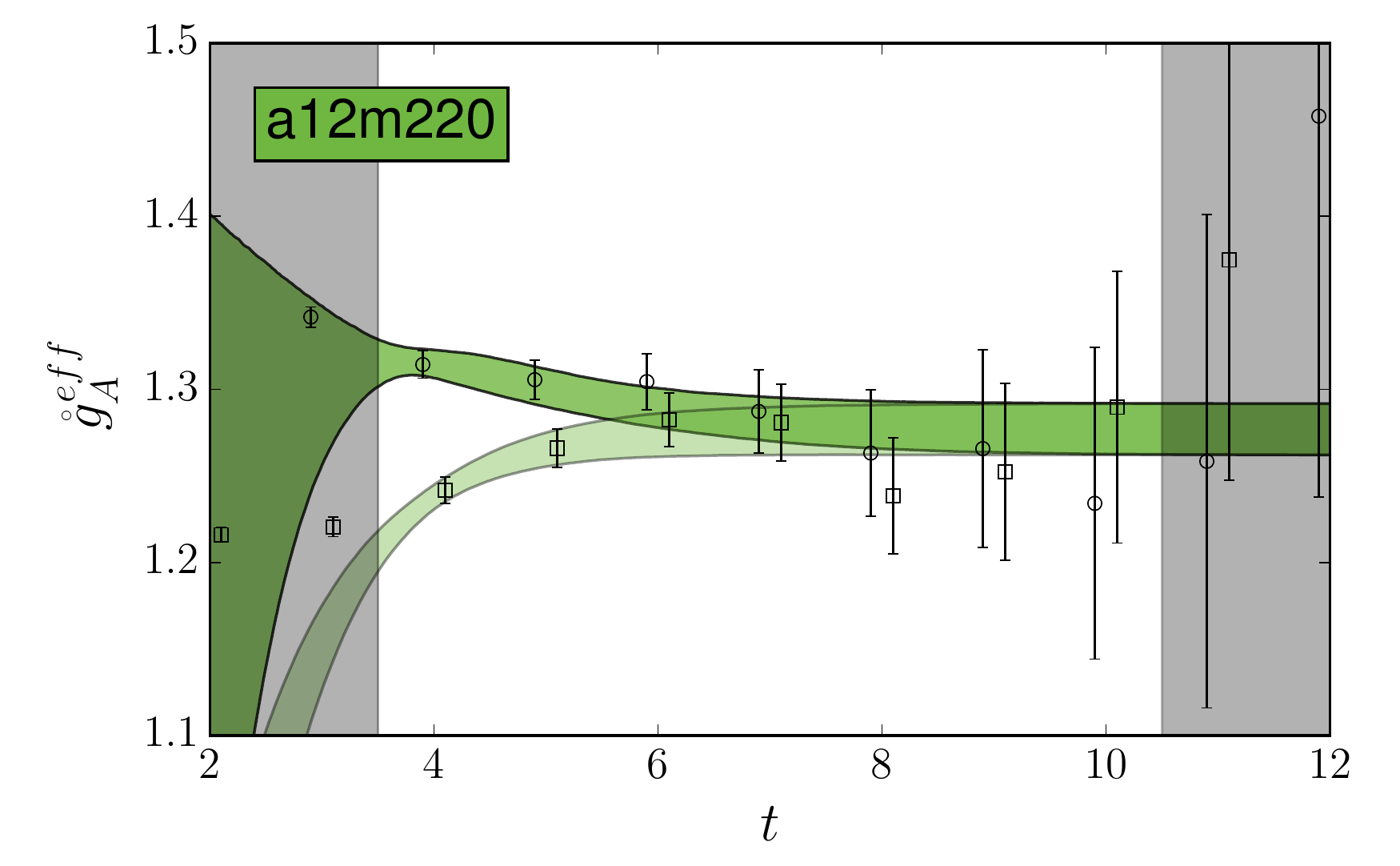}
	\includegraphics[width=0.5\columnwidth]{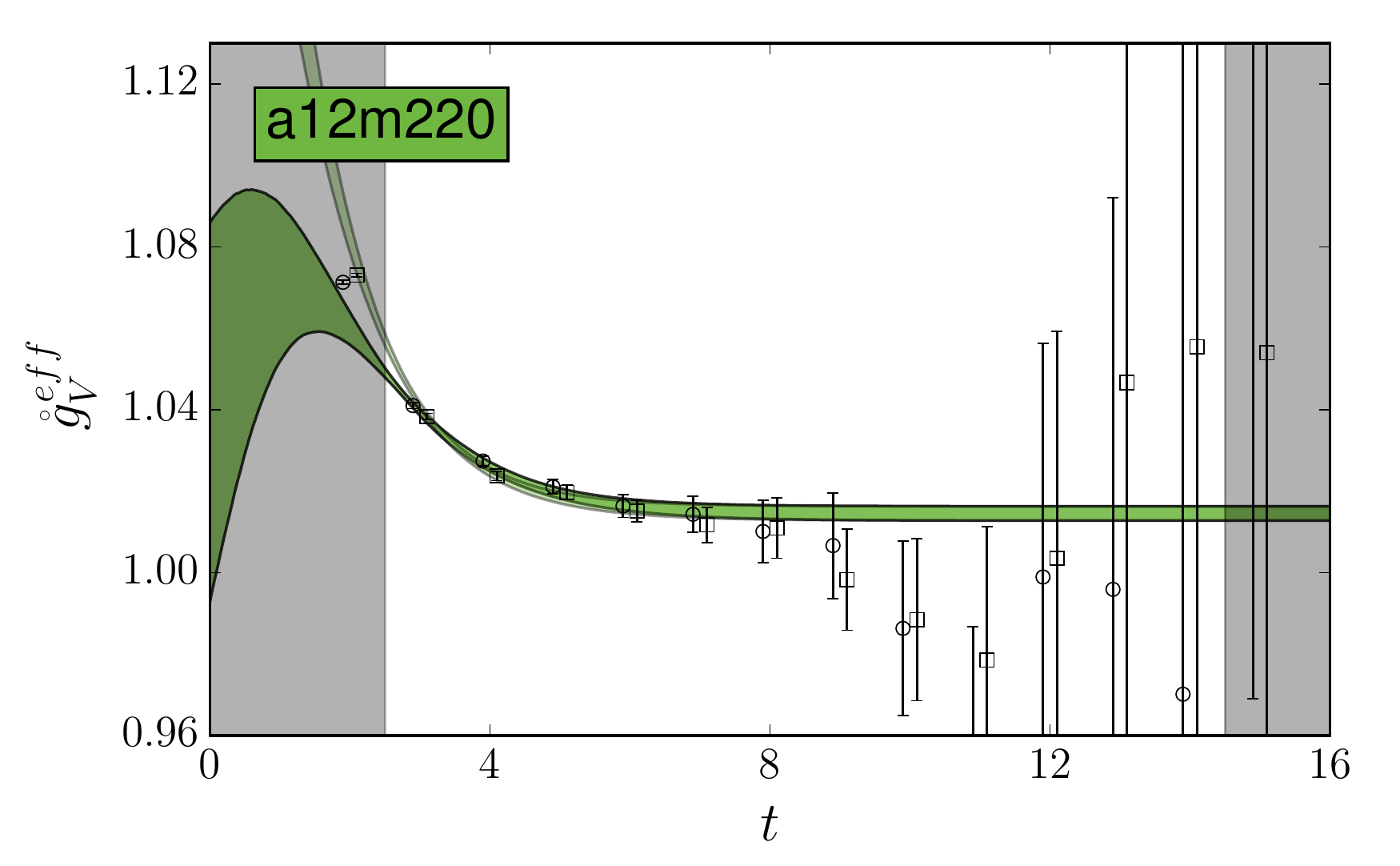}
	\caption{\label{fig:bs_example} Bootstrap distributions for the a12m220 ensemble with 5000 resamples. (Top Left) Bootstrap distribution  of the ratio of bare couplings. The shaded regions correspond to the middle 68\% and 95\% confidence intervals. (Top Right, Bottom Left, Bottom Right) The two-point effective mass, $g_A$ ratio, $g_V$ ratio for smeared (circle) and point (square) sink correlators. The green bands show the result of the preferred fit. The width of the green bands reflects the middle 68\% confidence interval. The grayed out regions indicate data that are not included in the preferred fit.}
\end{figure}

We propagate the resulting uncertainty of the correlator fits though bootstrap resampling. The simultaneous fit already accounts for correlations between the bare couplings, however we also account for correlations between the couplings and the pion mass and pion decay constants, which become relevant when the extrapolation to the physical point is performed. We take the central value of our preferred fit as the initial guess for the bootstrap routine, and sample the final distribution with 5000 bootstrap draws. We save a master list of random integer values for every ensemble, in which we coordinate all bootstrap routines for all calculations performed on these gradient-flowed HISQ ensembles. The bootstrap list and complete bootstrapped correlator results are saved in the collaboration PostgreSQL database hosted at NERSC. From the bootstrap distribution, we plot the histogram of the ratio of couplings, and the resulting fit on top of the correlator data. Fig.~\ref{fig:bs_example} provides an example of this check for the a12m220 ensemble. We observe that the correlator fit is numerically stable under bootstrap, showing no outliers and is clearly Gaussian distributed as expected. Plotting the fit result on top of the correlator data also reveals that the fit ansatz describes the data very well, capturing the curvature at early source-sink separation times, indicating that excited states are well described by a 2 state fit. Here we note that the nucleon suffers from a much larger signal-to-noise problem when compared to mesons, including heavy-light systems. This is reflected in the observation of correlated fluctuations which are observed at moderate values of source-sink separation time. On the other hand, when one studies heavy-light systems, while the signal damps away faster than for nucleon systems, the noise is also damped by a mass dominated by the bottom quark scale, which yields a noisy, but well-behaved uncertainty in the correlator. For the example shown in Fig.~\ref{fig:bs_example}, the correlated fluctuation can be seen in the bottom right panel. We observe that around 1~fm, the $g_V$ correlator exhibits a downward fluctuation of approximately one standard deviation. Therefore, aside from generating exponentially more data which will with brute force push the fluctuation to larger values of $t$, a controlled correlator fit to nucleon quantities can only be performed if one has access to small values of $t$. This is regularly performed for the two-point correlator, but will be more expensive if a fixed-sink method is used to construct three-point correlators.

\subsection{Pseudoscalar mass and decay constant fits}
The chiral-continuum extrapolation can be reparameterized to depend on the dimensionless quantity
\begin{align}
\epsilon_\pi\equiv \frac{m_\pi}{4\pi F_\pi}
\end{align}
circumventing the necessity of performing a scale-setting analysis.
Calculation of the pion mass, $m_\pi$, and the pion decay constant, $F_\pi$, are performed on the same lattice actions, gauge configurations, and sources as the main analysis of this paper.

A Bayesian constrained fit with a four-state fit ansatz is performed on the pion two-point correlation function in order to extract $m_\pi$ and its overlap factors; in Ref.~\cite{Berkowitz:2017opd}, we show that the oscillating state present in the Domain-Wall action are highly suppressed when the gauge fields are smeared with gradient-flow, and therefore are neglected in this analysis. A simultaneous fit to both the point-sink and smeared-sink correlators is performed, and statistics is further doubled by ``folding'' the meson correlators. Similarly, a Bayesian constrained fit to a constant is performed to extract $m_\text{res}$ from the $m_{\text{res}}$ correlator, as defined by Eq.~(5) of Ref.~\cite{Berkowitz:2017opd}. The 5D Ward Identity is used to obtain $F_\pi$ from $m_\pi$ and $m_{\text{res}}$ as given in Eq.~(6) of Ref.~\cite{Berkowitz:2017opd}.

The ground state priors for the pion mass and overlap factors are determined by the long-time limit of the effective mass and scaled correlators. The scaled correlator is the raw correlator with the leading exponential scaled away, and in the long time limit is proportional to the overlap factor. The ground state prior widths are set to 10\% of the prior central value, approximately two orders of magnitude larger than the width of the posterior distribution, thus leaving the ground state effectively unconstrained. The prior for the excited state energy splitting is log-normal and is set by approximately the mass splitting between the three-pion and one-pion state, with a width encompassing the two-pion to one-pion splitting within one standard deviation. The prior for $m_{\text{res}}$ is set by plotting the $m_{\text{res}}$ correlator, with a width set to 10\% of the central value, and is approximately one order of magnitude larger than its posterior distribution.

The fit regions for the pion two-point correlation functions and the $m_{\text{res}}$ correlators are chosen in the region of stability under varying choices of source-sink separation time to ensure full control over excited state systematic uncertainty.

 \begin{figure}
	\includegraphics[width=0.5\columnwidth]{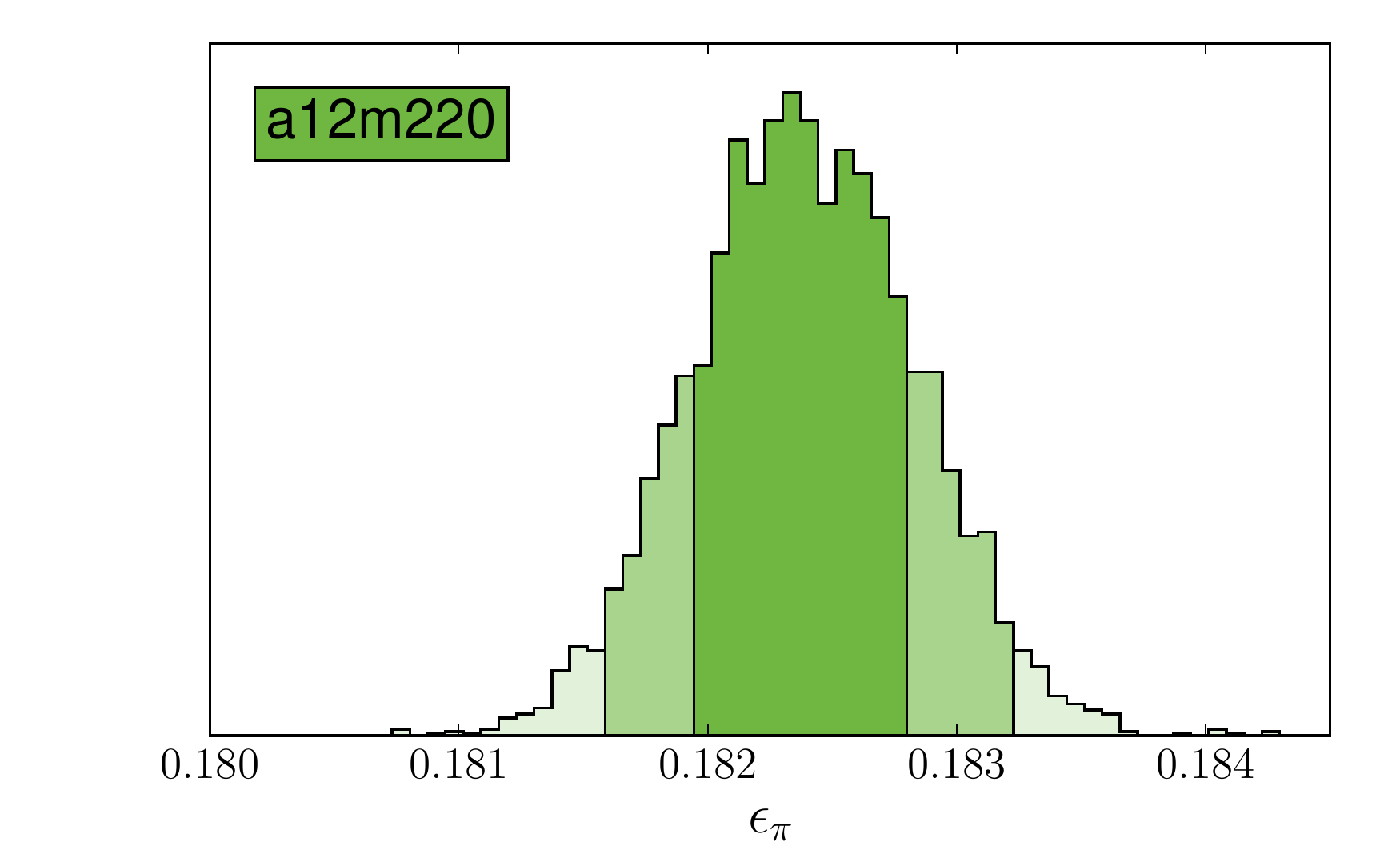}
	\includegraphics[width=0.5\columnwidth]{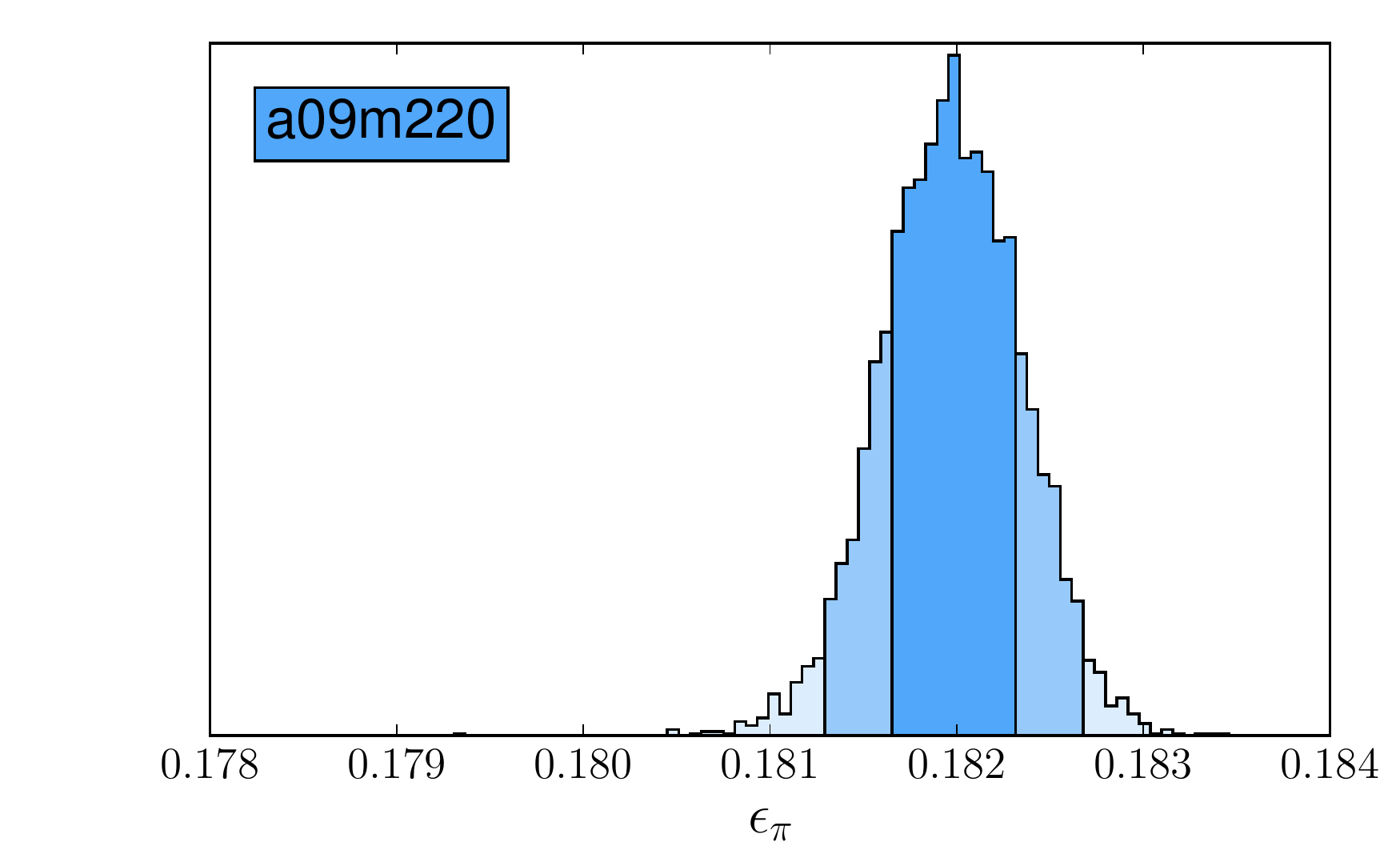}
	\caption{\label{fig:epi_example} Histograms of $\epsilon_\pi$ for the (Left) a12m220 and (Right) a09m220 ensembles with 5000 resamples. The shaded regions correspond to the middle 68\% and 95\% confidence intervals. The bootstraps are correlated with the histograms shown in Fig.~\ref{fig:bs_example}, facilitating a correlated analysis at later stages of the analysis.}
\end{figure}

Uncertainties are propagated by bootstrap resampling. For the preferred fits, 5000 bootstrapped configurations are analyzed, allowing for a correlated analysis with the bootstrapped samples of $\mathring{g}_A/\mathring{g}_V$. For each bootstrap, the prior central values for all parameters are set to a value randomly drawn from their corresponding initial prior distributions. The bootstrap histogram for $\epsilon_\pi$ for the a12m220 and a09m220 ensembles is provided in Fig.~\ref{fig:epi_example} as an example. We observe that the distributions are Gaussian without outliers, indicating that the constrained fits are numerically stable.

\section{Non-perturbative renormalization}
\label{sec5}
The discretization of the Dirac action leads to differences between the local current, as defined by Eq.~(\ref{eq:current_A},\ref{eq:current_V}), and the conserved currents. We correct for these differences using the non-perturbative Rome-Southampton renormalization procedure~\cite{Martinelli:1994ty}, with non-exceptional kinematics~\cite{Aoki:2007xm,Sturm:2009kb}, and implement momentum source quark propagators~\cite{Gockeler:1998ye}, achieving high statistical accuracy.

Defining the incoming and outgoing momentum-space propagators as $G(p)$ and $\bar{G}(p)$, and the quark bilinear matrix element $V_\Gamma(p_2,p_1)$, the amputated vertex function follows,
\begin{equation}
\Pi_\Gamma = \langle \bar{G}(p_2)^{-1}\rangle \langle V_\Gamma(p_2,p_1)\rangle \langle G(p_1)^{-1}\rangle.
\end{equation}

We match the quark bilinear matrix element to its tree-level value, indicated by the subscript `0', at scale $\mu$,
\begin{equation}
\left. Z_\Gamma \langle p_2 | \mathcal{J}_\Gamma(p_2,p_1) | p_1 \rangle \right|_{p_2^2=p_1^2=\mu^2} =\left.\langle p_2  | \mathcal{J}_\Gamma(p_2,p_1) | p_1 \rangle_0\right|_{p_2^2=p_1^2=\mu^2},
\end{equation}
where the quark bilinear operators $\mathcal{J}_\Gamma$ relevant to $g_A$ are defined by Eq.~(\ref{eq:current_A},\ref{eq:current_V}), and $Z_\Gamma$ is the renormalization coefficient. In particular, we enforce the SMOM condition such that $\mu^2 = p_2^2 = p_1^2 = (p_2-p_1)^2$. The matching to the tree-level value is performed in color and Dirac spaces, as a result, taking the trace yields the renormalization condition,
\begin{align}
Z_\Gamma \frac{\Lambda_\Gamma^{(s)}}{Z_q^{(s)}(\mu)} = F_\Gamma^{(s)}(\mu),
\end{align}
where $F_\Gamma^{(s)}$ is the trace of the tree-level matrix element, and $\Lambda_\Gamma^{(s)}$ is the amputated vertex function projected to the $s=\{\gamma_\mu,\slashed{q}\}$ intermediate schemes with $q=p_2-p_1$,
\begin{align}
	\Lambda_{\gamma_\mu}^{(\gamma_\mu)} = & \textrm{Tr}\left[ \Gamma \Pi_\Gamma\right],\\
	\Lambda_{\gamma_\mu}^{(\slashed{q})} = & \frac{q_\mu}{q^2} \textrm{Tr}\left[\slashed{q}\Pi_\Gamma\right].
\end{align}
We are only interested in the cases where $\Gamma=\{\gamma_\mu,\gamma_\mu\gamma_5\}$, in which case the intermediate schemes and choice of scale changes only the definition of the wave function renormalization factor $Z_q$ because the vector and axial currents are protected by Ward identities. We circumvent the calculation of $Z_q$ by taking advantage of the fact that the vector charge is by definition, normalized
\begin{equation}
Z_V \mathring{g}_V=1,
\end{equation}
and as a result, $g_A$ can be renormalized by computing the following ratio
\begin{equation}
g_A = \frac{Z_A}{Z_V}\frac{\mathring{g}_A}{\mathring{g}_V}.
\end{equation}
Therefore, in practice we compute the ratio of renormalization coefficients from
\begin{equation}
\frac{Z_A}{Z_V} = \frac{\Lambda_{\gamma_\mu}^{(s)}}{\Lambda_{\gamma_\mu\gamma_5}^{(s)}},
\end{equation}
which should be intermediate-scheme and scale independent.

\begin{figure}
	\includegraphics[width=0.5\columnwidth]{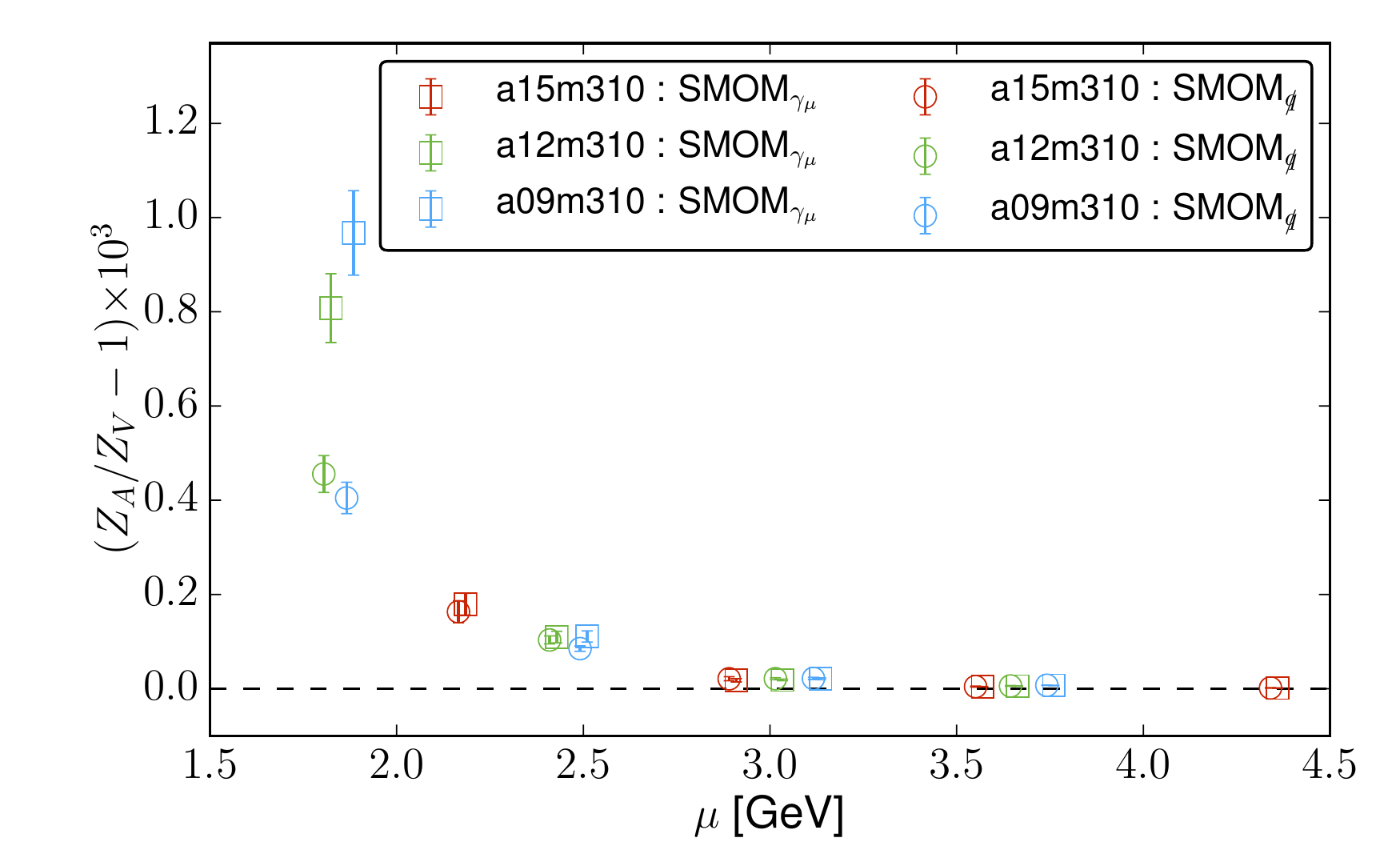}
	\includegraphics[width=0.5\columnwidth]{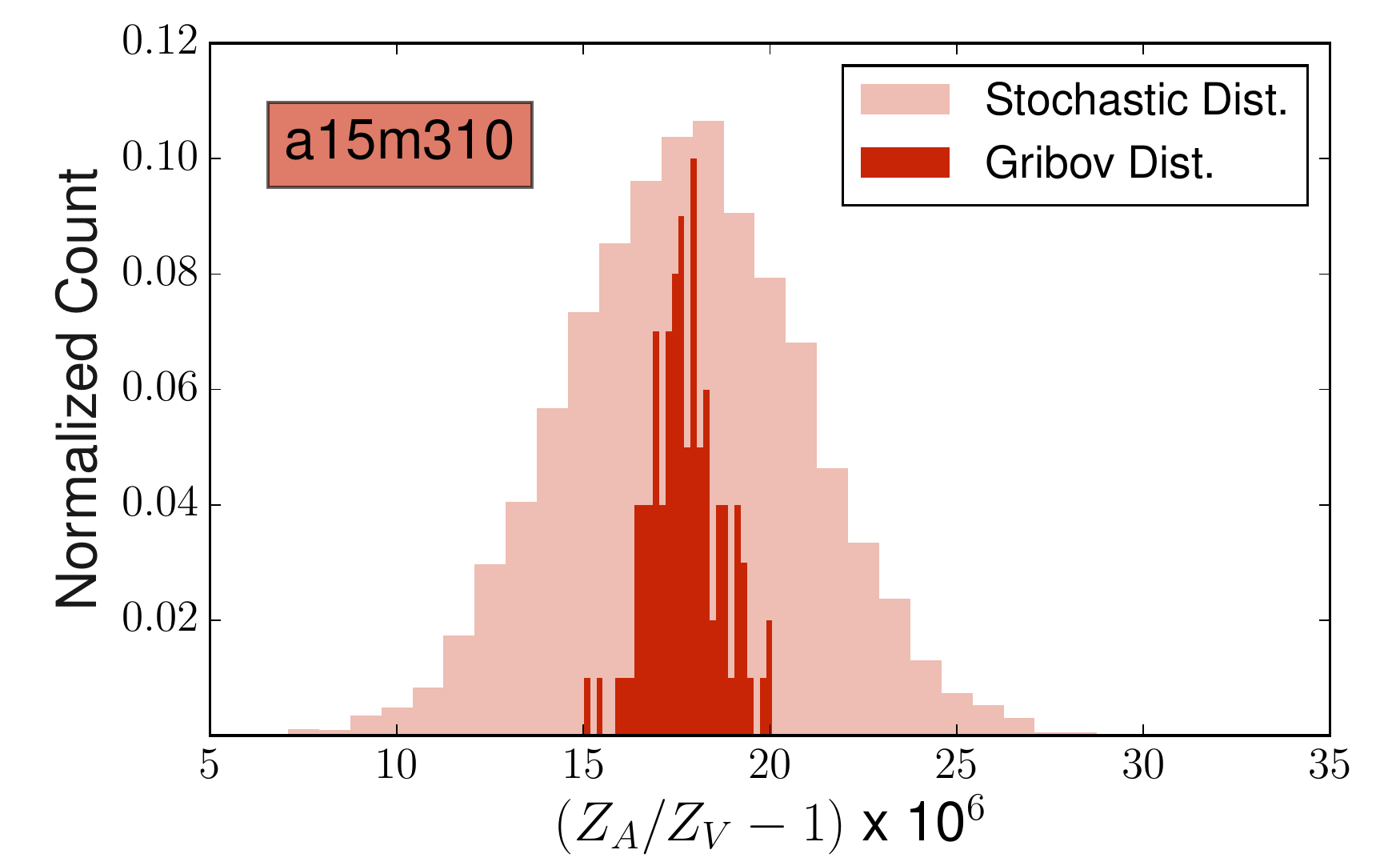}
	\caption{\label{fig:renorm} (Left) Renormalization coefficients for the $a=\{0.15,0.12,0.09\}$ ensembles with $m_\pi \sim 310$~MeV for both the $\gamma_\mu$ (square) and $\slashed{q}$ (circle) intermediate schemes. Note that the y-axis labels the deviation from unity at the per mille level. (Right) The Gribov distribution for the renormalization coefficient for the a15m310 ensemble plot against its statistical uncertainty at $\mu\sim 3$~GeV.}
\end{figure}

Fig.~\ref{fig:renorm} (Left) shows a calculation of the renormalization coefficients for the $a\sim\{0.15,0.12,0.09\}$ ensembles with $m_\pi\sim 310$~MeV. We observe that $Z_A$ and $Z_V$ from the $\gamma_\mu$ and $\slashed{q}$ schemes agree well within uncertainty above 2~GeV. Additionally, the ratio of coefficients enjoys a large Rome-Southampton window. Evidence of IR contamination vanishes for momenta larger than 3.0~GeV, while there is no evidence of an onset of growing discretization uncertainty for momenta through 4.5~GeV. Ratios of renormalization coefficients for all ensembles are commensurate with unity to one part in 10,000. Therefore, we do not quote the uncertainty related to these coefficients. Additionally, we study $Z_A/Z_V$ as a function of flow time and find the different to be insignificant.

Because quark-bilinear matrix elements are gauge variant, we perform these calculations under the Landau gauge. Landau gauge fixing however, is incomplete because different Gribov copies yields different local minima. We quantify this uncertainty by performing random global gauge transformations to the configurations before recalculating the renormalization coefficients. Fig.~\ref{fig:renorm} (Right) show this study on the a15m310 ensemble. We observe that the Gribov uncertainty is sub-dominant when compared to the statistical uncertainty.

\section{Chiral-continuum extrapolation}
\label{sec6}

\begin{figure}
	\includegraphics[width=0.5\columnwidth]{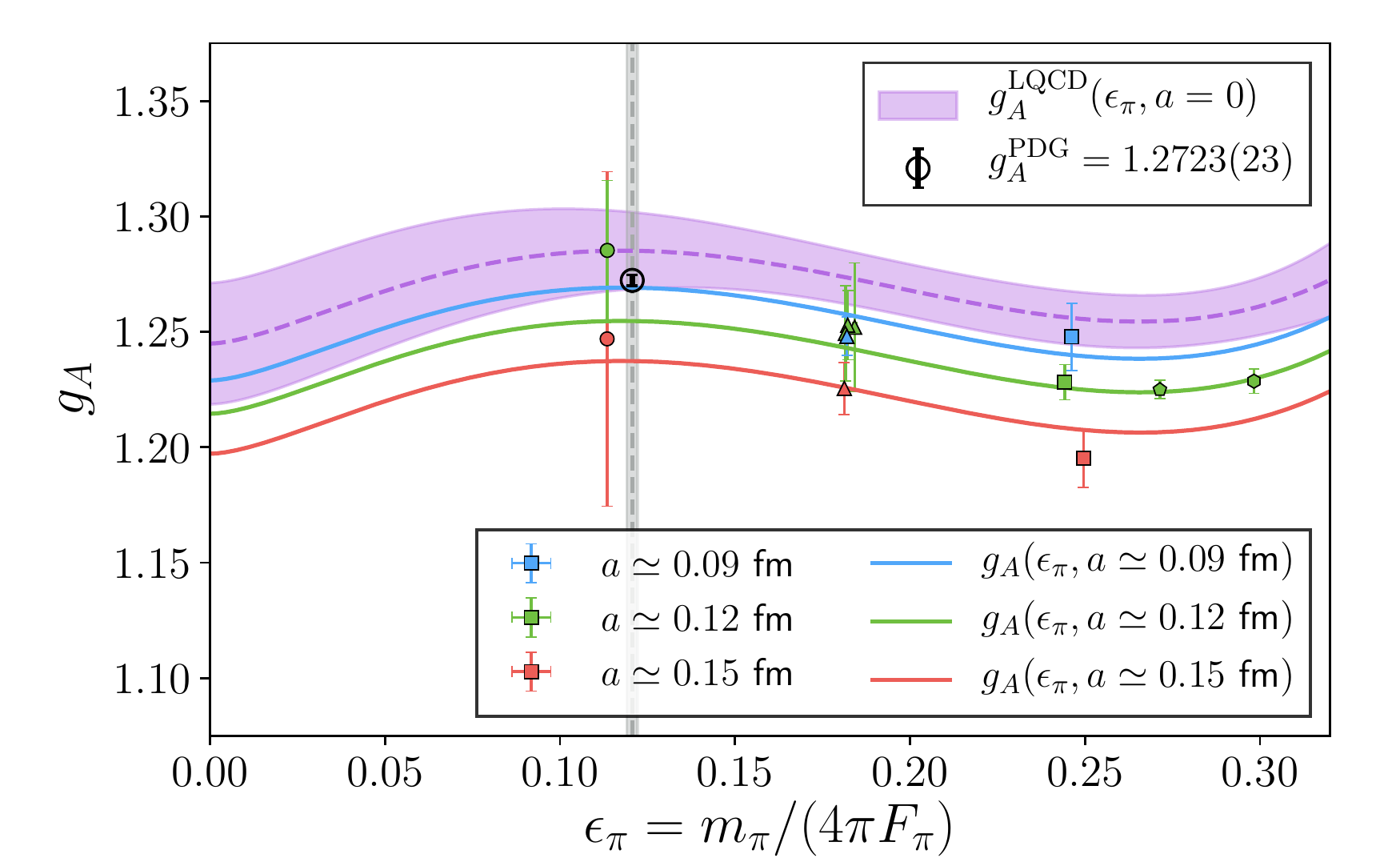}
	\includegraphics[width=0.5\columnwidth]{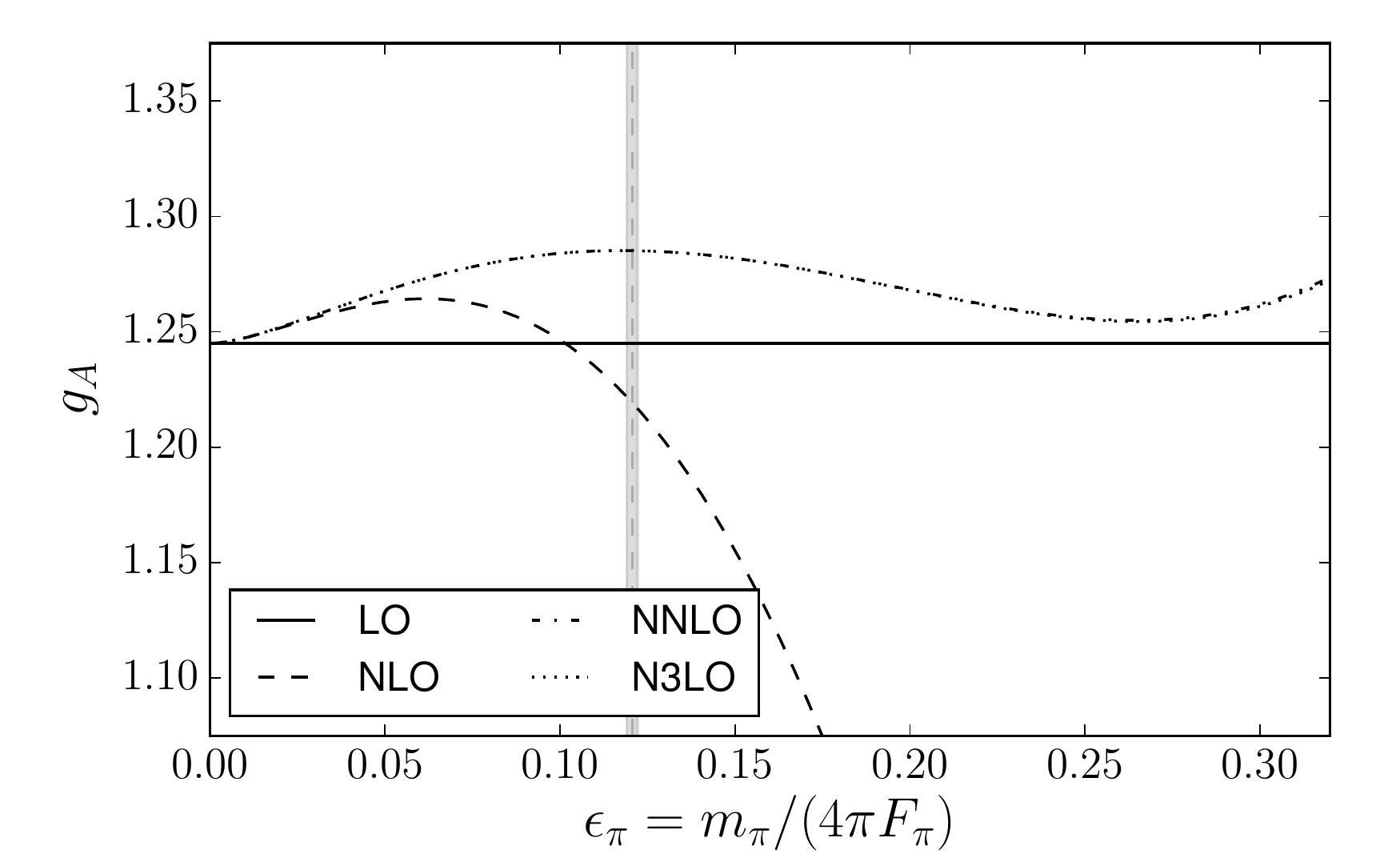}
	\includegraphics[width=0.5\columnwidth]{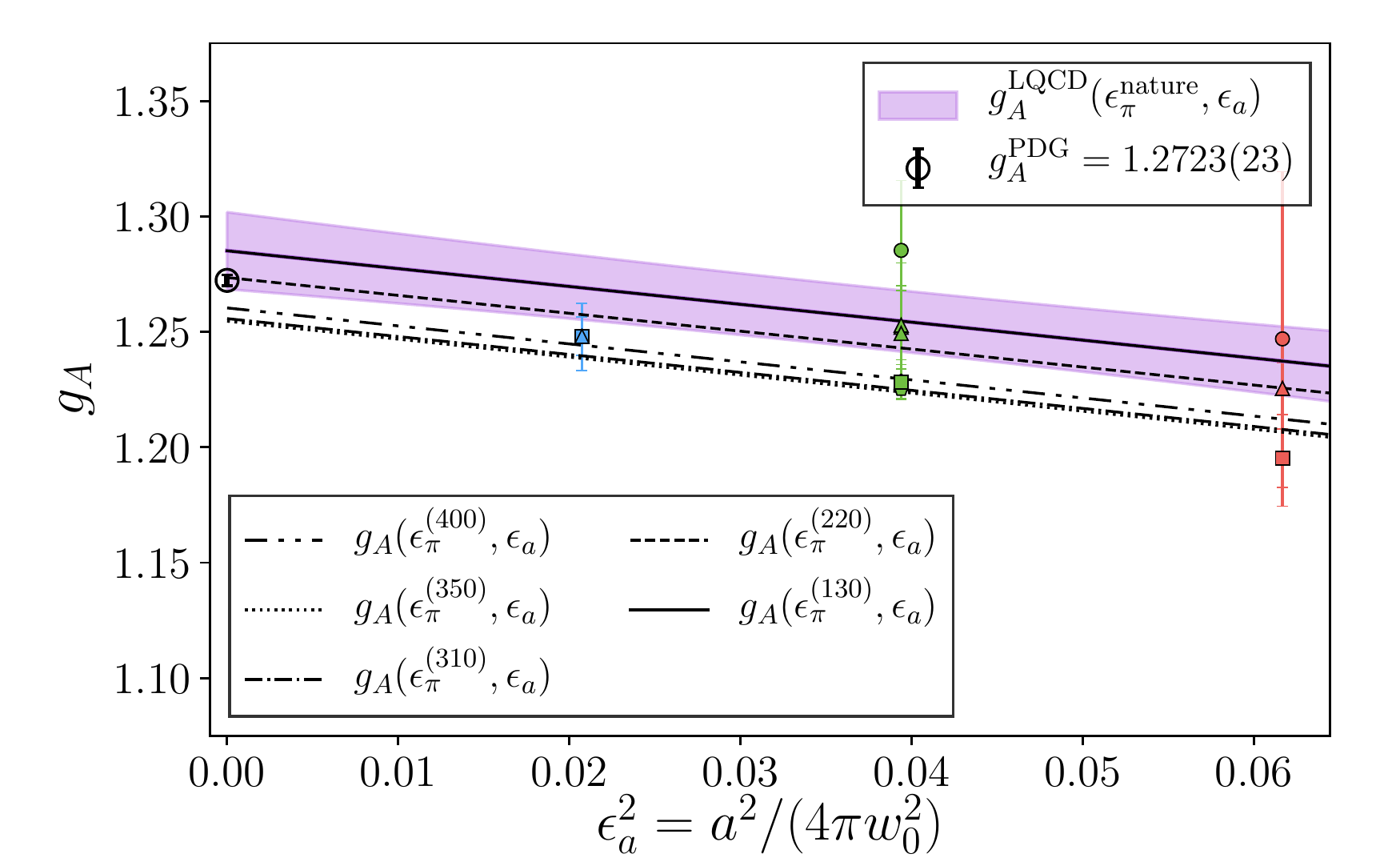}
	\includegraphics[width=0.5\columnwidth]{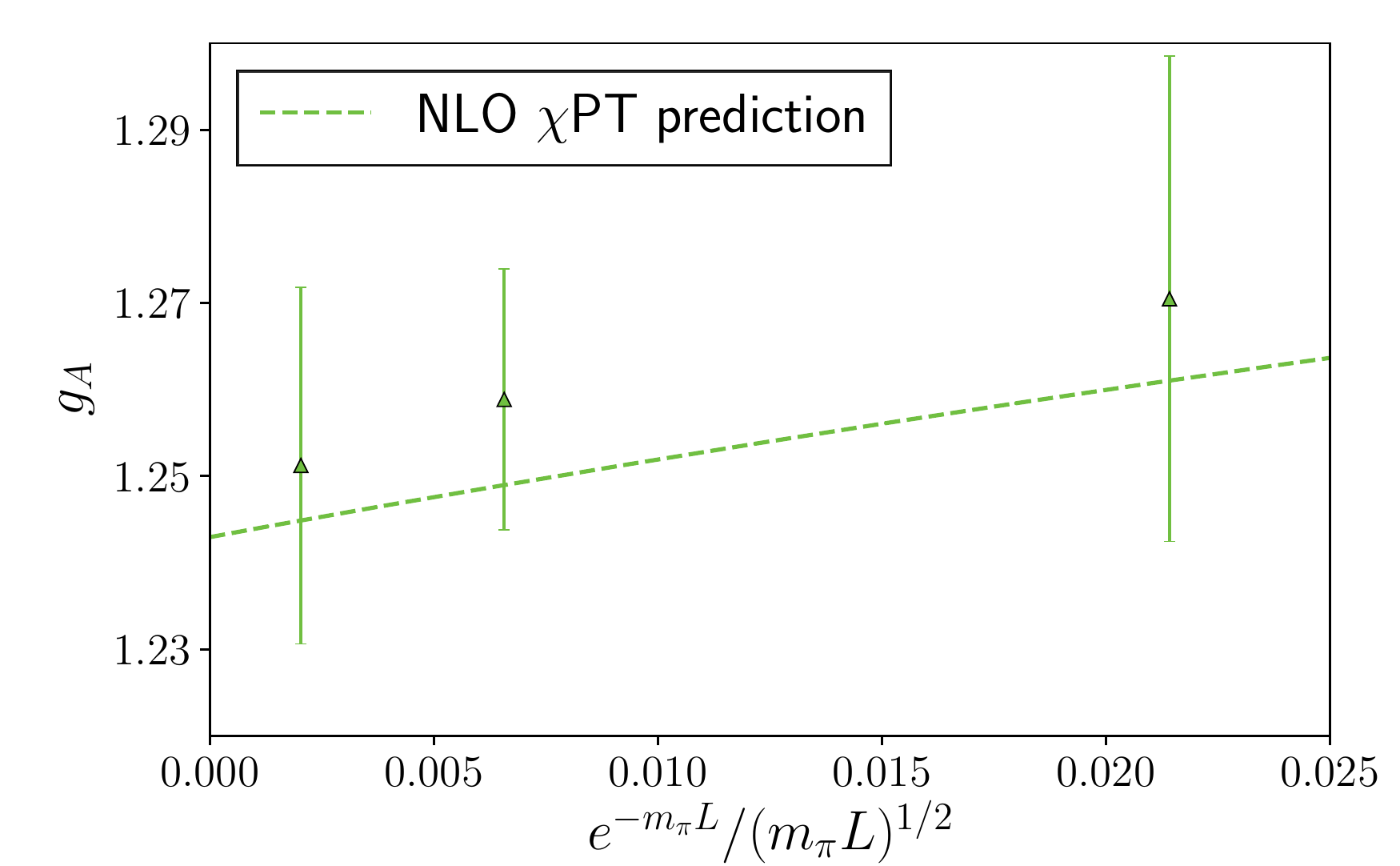}
	\caption{\label{fig:chipt} (Top Left) Chiral-continuum extrapolation of $g_A$ as a function of $\epsilon_\pi$. The red, green, and blue lines without error bars are the chiral extrapolation at finite lattice spacing, where $a = \{0.15, 0.12, 0.09\}$~fm respectively. The purple band is the continuum chiral-extrapolation, with propagated uncertainty. The vertical gray band indicates the physical $\epsilon_\pi$ value. The experimental value of $g_A$ is shown by the black circle marker. Pion masses for the lattice data are approximately 135, 220, 310, 350, and 400~MeV. The data and fit reconstruction are both extrapolated to infinite volume. (Top Right) The order-by-order convergence of the chiral extrapolation is presented. The LECs are determined by the preferred fit, before reconstructing the contribution from the $\chi$PT at each order. The vertical gray band indicates the physical $\epsilon_\pi$ value. (Bottom Left) Physical point extrapolation of $g_A$ as a function of $\epsilon_a^2$. The various black lines plot the continuum extrapolation at fixed pion masses without error bars, while the purple band plots the continuum extrapolation at the physical point mass. The experimental value of $g_A$ is indicated by the black circle marker. The data and fit reconstruction are both extrapolated to infinite volume.  (Bottom Right) The infinite volume extrapolation ($e^{-m_\pi L} / (m_\pi L)^{1/2} \rightarrow 0$) at $m_\pi \sim 220$~MeV is shown. The three greens points are the renormalized lattice matrix elements, and the green band is the resulting volume dependence prediction. Neither the points nor the fit reconstruction are extrapolated to infinite volume in this plot.}
\end{figure}

The renormalized lattice results for the axial coupling is extrapolated to the physical point using SU(2) NNLO Heavy Baryon chiral perturbation theory (HB$\chi$PT)~\cite{Jenkins:1990jv}. Since the axial coupling is a dimensionless quantity, the $\chi$PT is parameterized in terms of dimensionless parameters,
\begin{align}
\epsilon_\pi =& \frac{m_\pi}{4\pi F_\pi},\\
\epsilon_a^2 = & \frac{1}{4\pi}\left(\frac{a}{w_0}\right)^2,
\end{align}
such that the low-energy constants (LEC) are naturally $\mathrm{O}(1)$. The normalized pion decay constant is chosen such that $F_\pi\sim92$~MeV. The lattice spacing $a$ is made dimensionless by the gradient-flow scale $w_0$, which is precisely and accurately derived from Tab. IV of Ref.~\cite{Bazavov:2015yea}. The factor of $1/4\pi$ in the $\epsilon_a^2$ parameterization makes the numerical value of $\epsilon_\pi^2\sim\epsilon_a^2$, allowing for a double expansion around lattice spacing and pion mass with compatible power counting. We also observe that the $\epsilon_a$ parameterization results in $\mathrm{O}(1)$ coefficients for the leading discretization corrections.

The NNLO chiral extrapolation for $g_A$~\cite{Kambor:1998pi} is expressed as,
\begin{align}
g_A^{\chi\textrm{PT}} = & g_0 - \epsilon_\pi^2 \left[ (g_0 + 2g_0^3)\ln\left(\epsilon_\pi^2\right)-c_2\right] +  g_0 c_3 \epsilon_\pi^3.
\label{eq:chipt}
\end{align}
Additionally we include the $m_\pi^4$ analytic terms and up to NNLO discretization effects in the Symanzik expansion,
\begin{align}
g_A^{\textrm{analytic}} =& a_2 \epsilon_a^2 + c_4 \epsilon_\pi^4 + b_4 \epsilon_a^2 \epsilon_\pi^2 + a_4 \epsilon_a^4,
\end{align}
and NLO finite volume corrections~\cite{Beane:2004rf}
\begin{align}
\delta_L(\epsilon_\pi,m_\pi L) \equiv & g_A(L) - g_A(\infty)\nonumber \\
= & \frac{8}{3} \epsilon_\pi^2 \left[ g_0^3 F_1(m_\pi L) + g_0 F_3(m_\pi L) \right],
\end{align}
where $F_1, F_3$ are related to Bessel functions of the second kind.

The extrapolations to the physical pion mass, continuum limit, and infinite volume are performed simultaneously with the preferred fit function,
\begin{equation}
g_A(\epsilon_\pi,\epsilon_a,m_\pi L) = g_A^{\chi\textrm{PT}}(\epsilon_\pi) + g_A^{\textrm{analytic}}(\epsilon_\pi,\epsilon_a) + \delta_L(\epsilon_\pi,m_\pi L).
\end{equation}

Additionally, we consider the possibility of residual discretization errors of the form $a_1 a/w_0$ where $a_1 = \mathrm{O}(m_{\textrm{res}})$, and generic one-loop discretization errors of the form $s_2 \alpha_s \epsilon_a^2$ where $s_2 = \mathrm{O}(1)$.

We also check for model dependency by considering a Taylor expansion around the chiral point,
\begin{equation}
g_A^{\textrm{Taylor}} = c_0 + c_2\epsilon_\pi^2 + c_4 \epsilon_\pi^4 + a_2 \epsilon_a^2 + b_4 \epsilon_a^2 \epsilon_\pi^2 + a_4 \epsilon_a^4.
\end{equation}

We present our final chiral-continuum extrapolation  $g_A$ in Fig.~\ref{fig:chipt} (Top Left), and observe that NNLO $\chi$PT describes the data well, with a $\chi^2_{\textrm{aug.}}/\textrm{d.o.f.} = 0.43$. The fits are performed under the Bayesian framework, with $\mathrm{O}(10)$ priors for all LECs except for terms that are quadratic in $\epsilon^2_{a,\pi}$, which are set with $\mathrm{O}(1)$ priors. We use the definition of the augmented-$\chi^2$, and the corresponding degrees-of-freedom counting introduced in Appendix~B of Ref.~\cite{Bazavov:2016nty}. The chiral expansion converges after including the $\epsilon_\pi^3$ non-analytic term, as suggested by Fig.~\ref{fig:chipt} (Top Right). The curvature present in the pion mass extrapolation can be attributed to the large cancellation between the $\epsilon_\pi^2$ and $\epsilon_\pi^3$ terms of Eq.~(\ref{eq:chipt}). A plot of the physical point extrapolation as a function of the lattice spacing is shown in the bottom left of Fig.~\ref{fig:chipt}. There is no evidence of lattice spacing dependence that is quadratic in $\epsilon_a^2$. In fact, we observe minimal lattice spacing dependence, as our coarsest lattice matrix elements are within 6\% of our final extrapolated result. The bottom right of Fig.~\ref{fig:chipt} show the infinite volume extrapolation for the $m_\pi\sim220$~MeV ensembles. We observe that the finite volume prediction from NLO $\chi$PT adequately describes our data, and in addition, the finite volume matrix elements are consistent with the infinite volume extrapolation, demonstrating that finite volume corrections are a small effect.

\section{Systematic error analysis}
\label{sec7}
\begin{figure}
	\includegraphics[width=0.5\columnwidth]{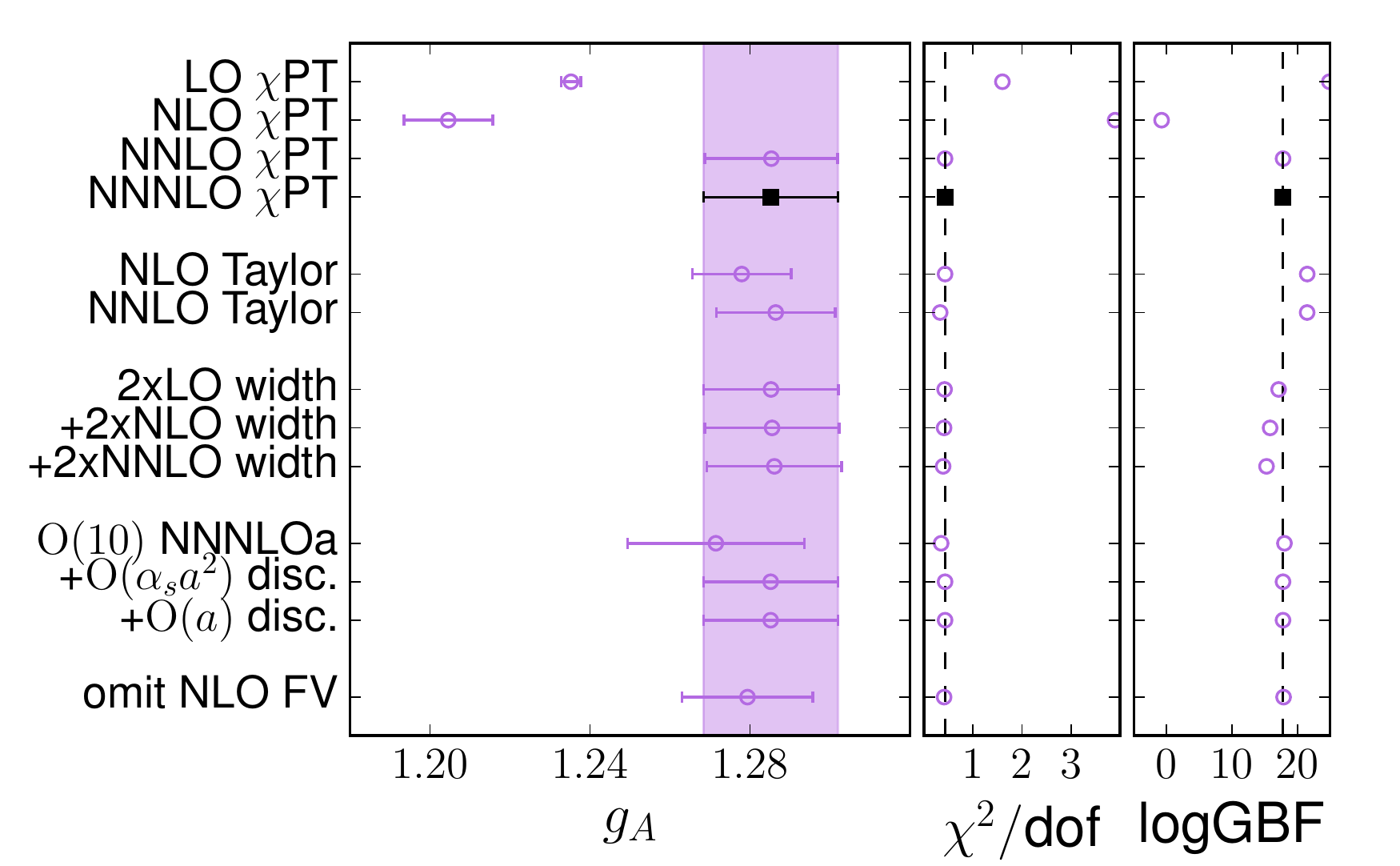}
	\includegraphics[width=0.5\columnwidth]{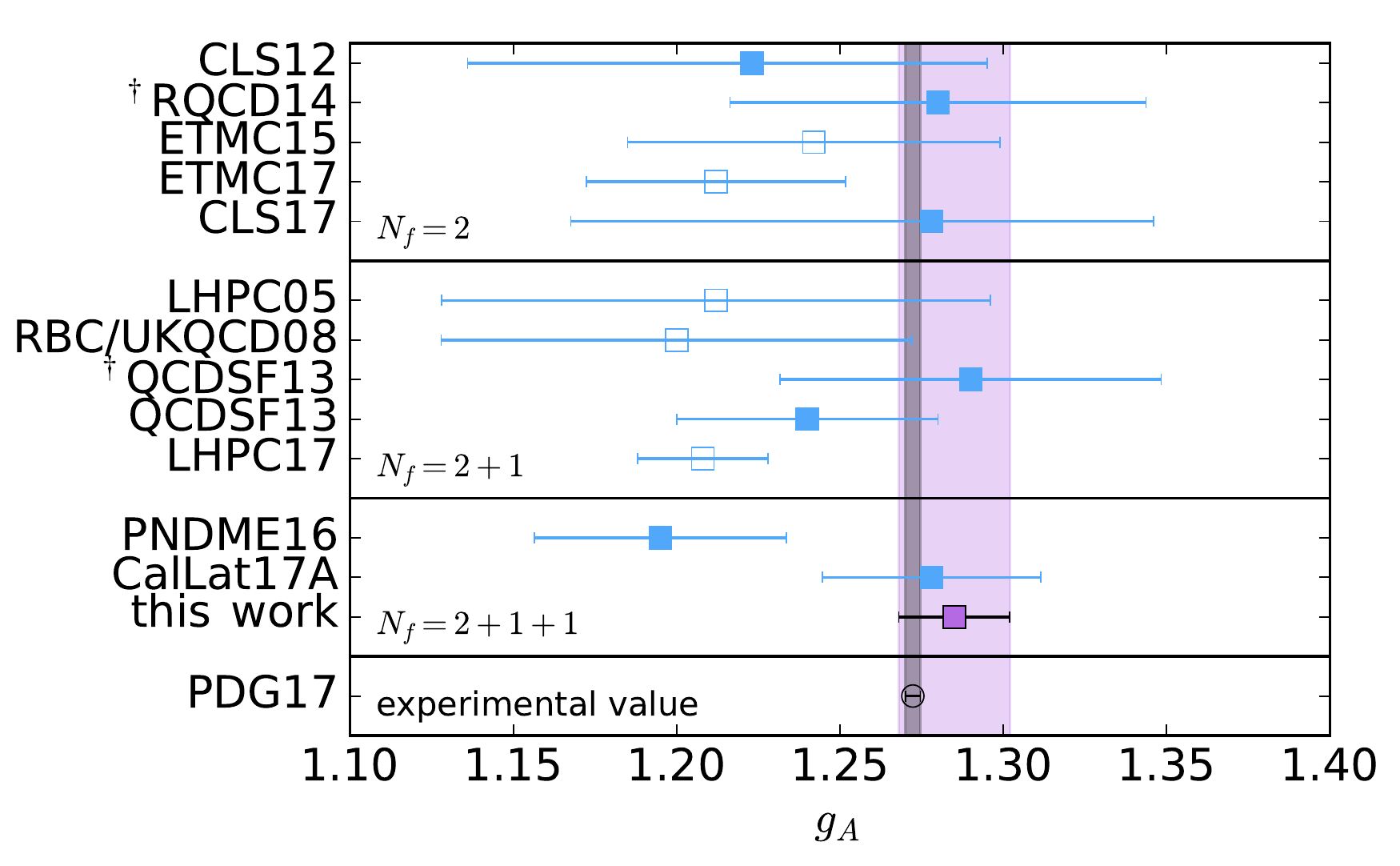}
	\caption{\label{fig:erroranalysis} (Left) Stability of the extrapolated value of $g_A$. From top to bottom, the first group of fits explores stability in $\chi$PT. The preferred fit lives within this group of fits and is highlighted in black. The next two fits are to a Taylor expansion, the next group labeled ``2x(N${}^n$)LO width'' test the dependency on priors, the next group evaluates the systematic uncertainty coming from the continuum extrapolation, and finally the systematic from truncating the finite volume correction at NLO is shown. The right two panels provides the $\chi^2_{\textrm{aug}}/\textrm{d.o.f.}$ and the log Bayes Factors for the various fits. The dashed vertical line in both panels highlights the value of the preferred fit. (Right) A summary of lattice results with regard to $g_A$ is compiled for LHPC05~\cite{Edwards:2005ym}, RBC/UKQCD08~\cite{Yamazaki:2008py}, CLS12~\cite{Capitani:2012gj}, QCDSF13~\cite{Horsley:2013ayv}, RQCD14~\cite{Bali:2014nma}, ETMC15~\cite{Abdel-Rehim:2015owa}, PNDME16~\cite{Bhattacharya:2016zcn}, LHPC17~\cite{Green:2017keo},CalLat17A~\cite{Berkowitz:2017gql}, ETMC17~\cite{Alexandrou:2017hac}, and CLS17~\cite{Capitani:2017qpc}. The averaged experimental determination is obtained from Ref.~\cite{Olive:2016xmw}. For simplicity, lattice results presented with statistical and systematic errors are added together in quadrature. The results with an open symbol are obtained from only one lattice spacing. The results with a $\dagger$ are extrapolated from the quantity $g_A/F_\pi$.}
\end{figure}

The robustness of the preferred chiral-continuum extrapolation is demonstrated by the stability of the extrapolated value of $g_A$ under an array of fit variations shown in Fig.~\ref{fig:erroranalysis}(Left). The group of fits labeled ``N${}^{\textrm{(n)}}$LO $\chi$PT'' shows that while the pion mass extrapolation receives large corrections up to NNLO, the fit stabilizes onwards.

The $\chi$PT extrapolation is checked against a linear and quadratic in $\epsilon_\pi^2$ Taylor expansion, and shows perfect consistency. Since the extrapolation is performed under a work, we check that priors of the LECs up to the $\epsilon_\pi^3$ non-analytic term are not constraining. Successively doubling the prior widths of the LECs result in insignificant changes in the final result.

The robustness of the continuum extrapolation is demonstrated in the following three fits: $\mathrm{O}(10)$~NNNLOa, $+\mathrm{O}(\alpha_s a^2)$~disc., $+\mathrm{O}(a)$~disc.. The first of the three fits increase the prior width of the $\epsilon_a^4$ LEC by one order of magnitude. Because the lattice calculation is performed only on three lattice spacings, we see that the posterior distribution of this LEC is determined mainly by the prior. We observe however, that the resulting fit is still consistent with the preferred fit, and assert that $\chi$PT power counting adequately captures the uncertainty at $\epsilon_a^4$. The next two fits in this group check for possible generic one-loop discretization errors, and the residual $\mathrm{O}(a)$ discretization error. We observe negligible dependence in both contributions.

The systematic uncertainty from the finite volume corrections are explored in the fit labeled ``omit NLO FV'', and the result is observed to be consistent with the preferred fit.

The right two panels of Fig.~\ref{fig:erroranalysis} (Right) show that with exception to the LO and NLO $\chi$PT fits, all fits describe the data well, with an $\chi^2_{\textrm{aug.}}/\textrm{d.o.f.}$ all around 0.5. On the other hand, the Bayes Factors show that if we ignore the Taylor expansion extrapolations, the preferred fit is the most likely model which reproduces the lattice correlator data. The $\chi$PT is physically motivated, in contrast to the Taylor expansion, and is the model used in the final result that is quoted.

Our post-diction of the nucleon axial coupling is
\begin{equation}
g_A^{\textrm{QCD}} = 1.285\pm 0.017
\end{equation}
with a final uncertainty budget of
\begin{table}[h]
\centering
\begin{tabular}{ll}
\textrm{statistical}                        & 1.29\% \\
\textrm{chiral extrapolation}         & 0.21\% \\
\textrm{continuum extrapolation} & 0.10\% \\
\textrm{infinite volume}                & 0.23\% \\
\textrm{isospin breaking}             & 0.04\% \\
\hline
\textrm{total}                               & 1.33\%
\end{tabular} \nonumber
\end{table}

Fig.~\ref{fig:erroranalysis} (Right) summarizes the improvement of the lattice determination of $g_A$ resulting from this work.

\section{Conclusions and Outlook}
\label{sec8}
The nucleon axial coupling has been a long standing challenge to determine using the methods of Lattice QCD. Two main challenges include overcoming the exponential noise problem, which in the case of the nucleon, leads to ill-behaved correlated fluctuations absent in meson quantities, as well as controlling the excited state contributions to the nucleon correlator. Using the Feynman-Hellmann method to generate lattice correlators, we are able to leverage data at small source-sink separation time in order to gain access to exponentially more precise data. In addition, the correlator stability plots demonstrate control over excited state contamination. As a result, we obtain a determination of $g_A$ at the percent-level commensurate with the experimental measurement. The post-diction of $g_A$ is the first step towards making a quantitative connection between nuclear physics and QCD, opening the door to calculations of nuclear quantities difficult or even inaccessible through experiment.

\section{Acknowledgments}
We thank C.~Bernard, A.~Bernstein, P.J.~Bickel, C.~Detar, A.X.~El-Khadra, W.~Haxton, V.~Koch, A.S.~Kronfeld, W.T.~Lee, G.P.~Lepage, E.~Mereghetti, G.~Miller, D.~Toussaint and F.~Yuan for discussions.
We thank the MILC Collaboration for providing their HISQ configurations~\cite{Bazavov:2010ru,Bazavov:2012xda} without restriction.
An award of computer time was provided by the Innovative and Novel Computational Impact on Theory and Experiment (INCITE) program to CalLat (2016) as well as the Lawrence Livermore National Laboratory (LLNL) Multiprogrammatic and Institutional Computing program through a Tier 1 Grand Challenge award.  This research used the NVIDIA GPU-accelerated Titan supercomputer at the Oak Ridge Leadership Computing Facility at the Oak Ridge National Laboratory, which is supported by the Office of Science of the U.S. Department of Energy under Contract No. DE-AC05-00OR22725, and the Surface and RZHasGPU clusters at LLNL.
This work was supported by the NVIDIA Corporation (MAC), the DFG and the NSFC Sino-German CRC110 (EB), an LLNL LDRD (EB, ER, PV), an LBNL LDRD (AWL), the RIKEN Special Postdoctoral Researcher Program (ER), the Leverhulme Trust (NG), the U.S. Department of Energy, Office of Science: Office of Nuclear Physics (EB, CMB, DAB, CCC, TK, HMC, AN, ER, BJ, KO, PV, AWL); Office of Advanced Scientific Computing (EB, TK, AWL); Nuclear Physics Double Beta Decay Topical Collaboration (DAB, HMC, AWL); and the DOE Early Career Award Program (DAB, CCC, HMC, AWL).
This work (EB, ER, PV) was performed under the auspices of the U.S. Department of Energy by LLNL under Contract No. DE-AC52-07NA27344.
Part of this work was performed at the Kavli Institute for Theoretical Physics supported by NSF Grant No. PHY-1125915.

\clearpage
\bibliography{lattice2017}

\end{document}